\documentclass[]{interact}

\usepackage{epstopdf}
\usepackage[caption=false]{subfig}

\usepackage[numbers,sort&compress]{natbib}
\bibpunct[, ]{[}{]}{,}{n}{,}{,}
\renewcommand\bibfont{\fontsize{10}{12}\selectfont}

\begin{document}

\articletype{PREPRINT}

\title{An iterative scheme for the generalized Peierls-Nabarro model based on the inverse Hilbert transform}

\author{
\name{Amuthan A. Ramabathiran\thanks{CONTACT Amuthan A. Ramabathiran. Email: amuthan@aero.iitb.ac.in}}
\affil{Department of Aerospace Engineering, Indian Institute of Technology Bombay, Mumbai 400076, India}
}

\maketitle

\begin{abstract}
A new semi-analytical iterative scheme is proposed in this work for solving the generalized Peierls-Nabarro model. The numerical method developed here exploits certain basic properties of the Hilbert transform to achieve the desired reduction of the non-local and non-linear equations characterizing the generalized Peierls-Nabarro model to a local fixed point iteration scheme. The method is validated with simple examples involving the 1D Peierls-Nabarro model corresponding to a sinusoidal stacking fault energy, and with calculations of the core structure of both edge and screw dislocations on the close-packed $\{111\}$ planes in Aluminium. An approximate technique to incorporate external stresses within the framework of the proposed iterative scheme is also discussed with applications to the equilibration of a dislocation dipole. Finally, the advantages, limitations and avenues for future extension of the proposed method are discussed. 
\end{abstract}


\section{Introduction}
It is well-known that the primary mediators for plasticity in crystalline materials are topological defects called dislocations. A proper understanding of the physics of dislocations, which is of crucial importance to many applications in materials science and engineering, entails considerations of processes from sub-atomic to continuum length and time scales. Notwithstanding the interplay of various factors across such an enormous range of scales, numerous simplified models that seek to explain specific aspects of dislocations have been proposed. One of the earliest, yet highly influential, models that throws light on the structure of dislocation cores is the Peierls-Nabarro (PN) model, originally proposed by Peierls \cite{peierls40}, and later elaborated upon by Nabarro \cite{nabarro47}. The current paper contributes a new iterative scheme to solve a particularly useful generalization of the original Peierls-Nabarro model of dislocations developed by Schoeck \cite{Schoeck94}. Specifically, the proposed method provides an efficient framework to model the slip distribution that characterizes the core structure of dislocations.

To set the context for the ensuing discussion, it is recalled that the energy of a crystalline solid containing a dislocation can be decomposed into two terms, one involving the elastic energy of the bulk, and another equal to the energy associated with the dislocation itself. Two common strategies are employed in the literature to effect this decomposition. The simpler of the two assigns an ad-hoc core radius to a dislocation and assumes linear elastic behavior outside the core region. The bulk energy associated with the region outside the defect core is computed using well established techniques in linear elasticity theory \cite{AHL16}. The energy of the core region, however, is not amenable to such a computation since the assumptions underlying linear elastic behavior are not valid in the dislocation core. A more serious limitation of the core radius approach is that the core structure of real dislocations is often complex enough (see, for instance, \cite{DR91}, \cite{RVCPW17}) that assigning a core radius fails to capture the essential features of the core structure. The PN model provides an alternative and more physically justified framework to study dislocation cores by taking advantage of the fact that dislocations are the boundaries between slipped and unslipped regions on well defined slip planes. The core structure is then determined by a competition between the elastic energy associated with the slip distribution and the energetic cost of the misfit between the slipped and unslipped regions. An attractive feature of the PN model is its variational structure, which permits a variety of systematic and controlled numerical approximations of the core structure \cite{KL72}, \cite{Lejcek76}, \cite{MGF98}, \cite{BK97}. In spite of the various assumptions inherent in the PN framework, it has been remarkably successful in explaining the core structure of dislocations in a variety of materials \cite{BK97}, \cite{LCWCS17}, \cite{Schoeck01b}, \cite{SHL18}, \cite{vSHW99}, \cite{LCWCS17b}.

There are many known limitations of the Peierls-Nabarro model of dislocations, two of which are briefly discussed here. First, the use of the generalized stacking fault energy as the misfit energy, as originally proposed by Vitek \cite{Vitek68}, \cite{CV70}, assumes a priori that the slip distribution varies slowly enough for the misfit energy to accurately represent the misfit energy. A second limitation of the PN model that is important for this work is that all information about the discreteness of the lattice plane on which the slip plane resides is lost, despite the fact that the generalized stacking fault energy is periodic with the periodicity of the lattice on the slip plane. Thus, there is no energetic cost to moving a dislocation on the slip plane according to the PN model. This is in sharp contrast to real dislocations which experience a lattice resistance to their motion due to the periodic energy landscape on the slip plane, called the Peierls potential. The stress required to overcome the Peierls potential, called the Peierls stress, is thus not captured by the PN model. An approximate means to estimate the Peierls stress was introduced by Nabarro \cite{nabarro47} by summing the total energy at discrete locations corresponding to the lattice points on the slip plane. This introduces a dependence of the total energy on the position of the dislocation line with respect to the atomic lattice, thereby recovering the lattice scale periodicity in the energy landscape that is absent in the original Peierls model. A succinct summary of other shortcomings of the PN model can be found in \cite{schoeck05}, \cite{Lu05}.

A lot of effort has gone into remedying the various shortcomings of the Peierls-Nabarro model: see, for instance, \cite{Schoeck94}, \cite{LCWCS17}, \cite{MPBO98}, for extensions of various aspects of the PN model, \cite{KCO02} for a modern reformulation of the Peierls-Nabarro model as a phase-field model, and \cite{wang15} for a discrete generalization of the PN model. Among these, the generalized Peierls-Nabarro model developed by Schoeck \cite{Schoeck94} is of note. The generalized PN model for an infinite dislocation assumes that the slip distribution has components both parallel and perpendicular to the dislocation line, and further uses anisotropic linear elasticity to model the elastic interactions. The generalized PN model has proven to be effective in providing insights into the core structure of dislocations in many materials; see, for instance, \cite{schoeck12}. The iterative solution scheme developed in this work is aimed at solving Schoeck's generalized PN model.

Among the many existing approaches to solve the generalized PN model, two classes of methods are widely used. The first class employs an ansatz for the slip distribution and/or the misfit stress distribution with the desired asymptotic properties and computes the slip distribution in a least squares sense. The ansatz for the slip distribution is chosen typically as a linear combination of the solution of the original PN equation. Instances of this approach can be found in \cite{KL72}, \cite{Lejcek76}, \cite{MGF98}, \cite{schoeck01}. An attractive feature of this approach is that the various parameters that enter the numerical model have straightforward physical interpretation. The second class of solution techniques is the so-called semi-discrete variational PN model (SVPN) that is based on a piecewise linear finite element discretization of the total energy of a crystalline solid with a dislocation \cite{BK97}. This is results in a nonlinear and nonlocal equation for the slip distribution which is solved using standard algorithms. It is to be noted that the process of solving these equations involves the use of dense matrices. Unlike the first class of methods mentioned earlier that assume a specific form of slip distribution, the approximation adopted in the SVPN method is ansatz-free. A variety of other methods have also been developed in the literature to solve the PN model: see \cite{WXM08} for a discrete Fourier transform based method to solve the generalized PN model for curved dislocations, and \cite{ZJZXH15} for a solution scheme based on the fast multipole method. The present contribution presents a novel computational scheme to solve the generalized PN model, and may be viewed as a competitive alternative to the aforementioned schemes.

The outline of the paper is as follows. Details regarding the mathematical formulation of the new iterative scheme for solving the generalized PN model are presented first. This is followed by a presentation of various examples that illustrate the applicability of the proposed model. Finally, the advantages, limitations and future extensions of the iterative method are discussed before concluding the paper. Relevant background information is summarized in two appendices.

\section{Mathematical formulation}
The mathematical formulation of a new iterative scheme to solve the generalized Peierls-Nabarro model is presented in this section. Towards this end, the generalized PN model is reviewed first. This is followed by a reformulation of the same to obtain an \emph{inverse Peierls-Nabarro model} by exploiting certain properties of the Hilbert transform; relevant properties of the Hilbert transform are summarize in Appendix~\ref{app:hilbert_transform}. The inverse PN model is subsequently approximated via a simple fixed point iteration scheme to obtain the desired iterative scheme that is the key contribution of this work. Unlike the generalized PN model, there is no straightforward means to incorporate the effect of an external stress within the framework of the inverse PN model; a possible strategy to include external stresses is finally discussed to complete the mathematical formulation of the newly proposed technique. 

\subsection{Recap of the generalized Peierls-Nabarro model}
\begin{figure}[h!]
\centering
\includegraphics[width=0.6\textwidth]{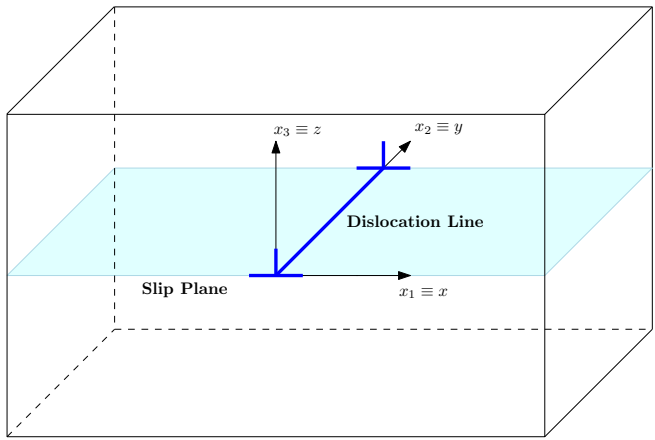}
\caption{Schematic illustration of the geometry of an infinite edge dislocation in an elastic crystalline material. An infinite screw dislocation has a similar geometry. The slip plane is assumed to be the $z=0$ plane, and the line direction of the dislocation is taken to be the $y$ axis. The Burgers vector of the dislocation is assumed to lie in the slip plane.}
\label{fig:dislocation_geometry}
\end{figure}

A summary of the generalized PN model introduced by Schoeck \cite{Schoeck94} is provided here, primarily for fixing the notation and terminology for the analysis to follow. The generalized PN model aims at describing the core structure of an infinite straight dislocation in an otherwise perfect crystal; a schematic illustration of an infinite (edge) dislocation is shown in Figure~\ref{fig:dislocation_geometry}.
The line direction of the dislocation is assumed to be along the $x_2$ direction\footnote{The notations $x, y, z$ are used interchangeably with $x_1, x_2, x_3$.}, along which it is infinite in extent. The glide plane, also referred to as the slip plane, of the dislocation is the $x_1x_2$ plane $x_3 = 0$. The Burgers vector of the dislocation is assumed to lie in the slip plane. The crystal is assumed to be infinite along all three directions. The total energy of the crystalline solid with a single dislocation is modeled as the sum of two contributions: the elastic energy of the two halves of the material separated by the slip plane, and the energy associated with the specific slip distribution on the slip plane. Owing to the geometry chosen for this problem, the slip distribution is assumed to be uniform along the $x_2$ direction. Furthermore, the slip is assumed to be entirely contained in the $x_1x_2$ plane. With these assumptions in place, the slip distribution on the slip plane is of the form $(s_1(x), s_2(x))$. The total energy of the dislocated crystal with the slip distribution $(s_1(x), s_2(x))$ can thus be written as\footnote{The following notation is used: greek indices take the values $1$ and $2$. The Einstein convention regarding repeated indices is \emph{not} followed; all summations are explicitly indicated, though the range of the variables is often to be inferred from the context.}
\begin{equation} \label{eq:pn_energy}
E(s_1, s_2) = \sum_{\alpha,\beta}\frac{K_{\alpha\beta}}{4\pi}\int_{-\infty}^{\infty}\int_{-\infty}^{\infty} \frac{\rho_\alpha(x') s_\beta(x)}{x - x'}\,dx\,dx' + \int_{-\infty}^{\infty} \gamma(s_1(x), s_2(x))\,dx.
\end{equation} 
Here,
\begin{equation} \label{eq:dislocation_density}
\rho_\alpha(x) = \frac{ds_\alpha(x)}{dx}
\end{equation}
denotes the component of the dislocation density along the $x_\alpha$ direction. The second term in \eqref{eq:pn_energy} represents the misfit energy associated with a slip distribution $(s_1(x), s_2(x))$ over the slip plane. A convenient choice for the misfit energy is obtained by choosing $\gamma(\delta_1, \delta_2)$ to be the the generalized stacking fault energy obtained as the elastic energy due to a uniform relative slip $(\delta_1, \delta_2)$ of the upper ($z > 0$) half of the crystal relative to the lower half \cite{CV70}. $\{K_{\alpha\beta}/4\pi\}$ in \eqref{eq:pn_energy} are the components of the symmetric anisotropic Stroh tensor \cite{BBS79} which depends on the crystalline lattice and the geometry of the slip plane. The equilibrium slip distribution is then obtained as the minimizer of the energy functional \eqref{eq:pn_energy}. This yields the generalized PN equation:
\begin{equation} \label{eq:gen_pn}
\sum_\beta \frac{K_{\alpha\beta}}{2\pi}\int_{-\infty}^{\infty} \frac{\rho_\beta(x')}{x - x'}\,dx' = \tau_\alpha(s_1(x), s_2(x)),
\end{equation}
where
\begin{equation} \label{eq:pn_misfit_stress}
\tau_\alpha(\delta_1, \delta_2) = -\partial_\alpha \gamma(\delta_1, \delta_2)
\end{equation}
is the stress associated with a uniform slip $(\delta_1, \delta_2)$, called henceforth the misfit stress. The slip distribution corresponding to the presence of a single dislocation is obtained by solving the generalized PN equation, \eqref{eq:gen_pn}, subject to the boundary conditions $s_1(-\infty) = 0, s_1(\infty) = b, s_2(\pm \infty) = 0$. 

The original PN model (\cite{peierls40}, \cite{nabarro47}) is obtained by considering an isotropic elastic medium with slip distribution constrained by the condition $s_2(x) \equiv 0$. Assuming a simple sinusoidal form of the stacking fault energy,
\begin{equation} \label{eq:gamma_sin_1d}
\gamma(\delta) = \frac{\mu b^2}{4\pi^2d}\left(1 - \cos \frac{2\pi \delta}{b}\right),
\end{equation}
the PN equation reduces to the following equation:
\begin{equation} \label{eq:pn_iso_1d}
\int_{-\infty}^{\infty} \frac{\rho(\xi)}{x - \xi}\,d\xi = -\frac{1 - \nu}{d}\sin \frac{2\pi s(x)}{b},
\end{equation}
where $s(x) = s_1(x)$, $\mu$ is the shear modulus and $\nu$ is the Poisson's ratio of the isotropic elastic medium. This equation, with the boundary conditions stated earlier admits an exact solution:
\begin{equation} \label{eq:pn_iso_1d_soln}
s(x) = \frac{b}{2} + \frac{b}{\pi} \tan^{-1} \frac{x}{\xi}, \quad \xi = \frac{d}{2(1 - \nu)}.
\end{equation}
This solution will be used as a basic benchmark for the numerical methods presented later. The quantity $\xi$ serves as a useful characterization of the width of the dislocation.

\subsection{Inverse Peierls-Nabarro model}
The generalized PN equation \eqref{eq:gen_pn} is now reformulated in a form that will prove to be useful in the sequel. Using the definition of the Hilbert transform (see Appendix~\ref{app:hilbert_transform}), the generalized PN equation \eqref{eq:gen_pn} can be written as
\begin{equation} \label{eq:pn_hilbert}
\mathcal{H}\left(\sum_\beta \frac{1}{2}K_{\alpha\beta}\rho_\beta(x)\right) = \tau_\alpha(s_1(x), s_2(x)).
\end{equation} 
The key idea behind the numerical strategy proposed in this work is the observation that if the misfit stress distribution $\tau_\alpha(s_1(x), s_2(x))$ can be expressed as the Hilbert transform of some function $g_\alpha(x)$, as in
\begin{displaymath}
\tau_\alpha(s_1(x), s_2(x)) = \mathcal{H}(g_\alpha(x)),
\end{displaymath}
then the generalized PN equation \eqref{eq:gen_pn} can be replaced by an equation of the form
\begin{displaymath}
\mathcal{H}\left(\sum_\beta \frac{1}{2} K_{\alpha\beta}\rho_\beta(x)\right) = \mathcal{H}(g_\alpha(x)) \quad\Rightarrow\quad \sum_\beta \frac{1}{2} K_{\alpha\beta}\rho_\beta(x) = g_\alpha(x) + l_\alpha,
\end{displaymath}
where $l_\alpha$ is a constant. Such an equation is much simpler to handle numerically compared to the generalized PN model \eqref{eq:gen_pn}, as will become evident shortly. This is accomplished through a sequence of steps as follows. To begin with, the periodicity of the lattice implies that the generalized stacking fault energy $\gamma(\delta_1, \delta_2)$ is a periodic function of both its arguments, with periodicity $b_1$ and $b_2$ along the $x_1$ and $x_2$ directions, respectively. It is thus possible to express the generalized stacking fault energy in terms of a Fourier series of the form
\begin{equation} \label{eq:gamma_fourier}
\gamma(\delta_1, \delta_2) = \sum_{m_1=-\infty}^{\infty} \sum_{m_2=-\infty}^{\infty} c_{m_1,m_2} \exp \left(i \sum_\beta m_\beta k_\beta \delta_\beta\right),
\end{equation}
where 
\begin{equation} \label{eq:def_k1k2}
k_\alpha = \frac{2\pi}{b_\alpha},
\end{equation}
and $\{c_{m_1,m_2}\}$ are complex constants. It is noted that in practice only a finite number of such coefficients $c_{m_1,m_2}$ are non-zero. A further reduction in the number of coefficients $c_{m_1,m_2}$ that need to be computed is achieved by noting that $\gamma(\delta_1,\delta_2)$ is real. Using \eqref{eq:pn_misfit_stress}, the misfit stresses $\tau_\alpha(\delta_1, \delta_2)$ are obtained from the Fourier expansion of the generalized stacking fault energy \eqref{eq:gamma_fourier} as
\begin{equation} \label{eq:tau_fourier}
\tau_\alpha(\delta_1, \delta_2) = -\sum_{m_1,m_2} i m_\alpha k_\alpha c_{m_1,m_2} \exp \left(i \sum_\beta m_\beta k_\beta \delta_\beta\right).
\end{equation}
Note that $\tau_\alpha(\delta_1, \delta_2)$ is also a periodic function with periodicities $b_1$ and $b_2$ along the $x_1$ and $x_2$ directions. However, given an arbitrary slip distribution $(s_1(x), s_2(x))$, the misfit stress distribution $\tau_\alpha(s_1(x), s_2(x))$ is in general not a periodic function of $x$. It is helpful to express the misfit stress distribution using Fourier transform as
\begin{equation} \label{eq:tau_slip_ft}
\begin{split}
\tau_\alpha(s_1(x), s_2(x)) &= \frac{1}{2\pi} \int_{-\infty}^{\infty} \hat{d}_\alpha(k) \exp (ikx) \, dk,\\
\hat{d}_\alpha(k) &= \int_{-\infty}^{\infty} \tau_\alpha(s_1(x), s_2(x)) \exp (-ikx) \, dx.
\end{split}
\end{equation}
Using \eqref{eq:ht_expi} and \eqref{eq:ht_inv}, it is easily seen that
\begin{equation} \label{eq:tau_hilbert_approx}
\tau_\alpha(s_1(x), s_2(x)) = \mathcal{H}\left(\frac{i}{2\pi}\int_{-\infty}^{\infty} \text{sgn}(k) \hat{d}_{\alpha}(k) \exp (ikx) \, dk\right).
\end{equation}
Since \eqref{eq:tau_hilbert_approx} expresses the misfit stress distribution as the Hilbert transform of some function, the arguments given in the beginning of this section can be used to invert the generalized Peierls-Nabarro model. Specifically, using \eqref{eq:tau_hilbert_approx} in \eqref{eq:pn_hilbert} and using the inverse Hilbert transform yields the following equation:
\begin{equation} \label{eq:inverse_pn_model}
\sum_\beta \frac{1}{2}K_{\alpha\beta}\rho_\beta(x) = \frac{i}{2\pi}\int_{-\infty}^{\infty} \text{sgn}(k) \hat{d}_{\alpha}(k) \exp (ikx) \, dk + l_\alpha,
\end{equation} 
where $l_\alpha$ is a constant determined by the boundary conditions. Since the process of deriving \eqref{eq:inverse_pn_model} from \eqref{eq:pn_hilbert} is effected with the aid of the inverse Hilbert transform, the foregoing equation \eqref{eq:inverse_pn_model} is henceforth referred to as the \emph{inverse Peierls-Nabarro model}. It is emphasized that the inverse PN model \eqref{eq:inverse_pn_model} is essentially exact, and it is easy to check that solutions of the inverse PN model are identical to those of the generalized PN model \eqref{eq:gen_pn}.

The inverse PN model \eqref{eq:inverse_pn_model} does not admit an analytical solution in general. Nevertheless, an analytically tractable approximation of the inverse PN model \eqref{eq:inverse_pn_model} can be obtained as follows. Since the lattice structure is almost perfectly restored far away from the dislocation, it is clear that $s_\alpha(L) = b_\alpha$ and $s_\alpha(-L) = 0$ asymptotically for sufficiently large $L$. This in turn implies that $\tau_\alpha(s_1(\pm L), s_2(\pm L)) \simeq 0$. Thus, if the domain of slip is modeled as $[-L,L]$ along the $x$ axis, for some sufficiently large $L$, then the Fourier transform \eqref{eq:tau_slip_ft} of $\tau_\alpha(s_1(x), s_2(x))$ can be approximated using a Fourier series of the form
\begin{equation} \label{eq:tau_slip_fourier}
\tau_\alpha(s_1(x), s_2(x)) \simeq \sum_{n = -\infty}^{\infty} d_{\alpha,n} \exp ink_0x,  
\end{equation}
where
\begin{equation} \label{eq:def_k0}
k_0 = \frac{\pi}{L},
\end{equation}
and the Fourier coefficients $d_{\alpha,n}$ in \eqref{eq:tau_slip_fourier} are related to the coefficients $d_\alpha(k)$ in \eqref{eq:tau_slip_ft} as
\begin{equation} \label{eq:tau_slip_ft_fourier_coeff}
d_{\alpha,n} = \frac{1}{2L}\hat{d}_{\alpha}(nk_0).
\end{equation}
In practice, the summation in \eqref{eq:tau_slip_fourier} is carried over a finite range $n \in [-N_\tau, N_\tau]$. To obtain analytical expressions for $\hat{d}_\alpha(k)$ (and thereby the coefficients $\{d_{\alpha,n}\}$), the slip distribution is further approximated as a piecewise linear function over a suitable discretization of the domain $[-L,L]$: for $j = 0, 1, \ldots, (N_s - 1)$, 
\begin{equation} \label{eq:piecewise_lin_slip}
s_\alpha(x) = s_{\alpha,j} + \frac{s_{\alpha, j + 1} - s_{\alpha, j}}{h_j}(x - x_j), \quad x \in [x_j, x_{j + 1}].
\end{equation}
Here $(x_0, \ldots, x_{N_s})$ is a partition of $[-L,L]$, with $x_0 = -L$ and $x_{N_s} = L$, $h_j = x_{j+1} - x_{j}$, and $s_{\alpha,i} = s_\alpha(x_i)$. Using the piecewise linear approximation \eqref{eq:piecewise_lin_slip} for the slip distribution, the Fourier expansion \eqref{eq:gamma_fourier} of the stacking fault energy, and the definition of the misfit stress \eqref{eq:pn_misfit_stress}, the coefficients $\hat{d}_\alpha(k)$ of the Fourier transform of $\tau_\alpha(s_1(x), s_2(x))$ in \eqref{eq:tau_slip_ft} can be computed analytically as follows:
\begin{equation} \label{eq:tau_slip_ft_coeff}
\begin{split}
\hat{d}_\alpha(k) &= -\frac{i}{2L}\sum_{j=0}^{N_s - 1}\sum_{m_1,m_2} m_\alpha k_\alpha c_{m_1,m_2} \exp \left(i \sum_\beta m_\beta k_\beta \lambda_{\beta,j} \right) I^n_{j,m_1,m_2},\\
I^n_{j,m_1,m_2} &= \begin{cases}
h_j, & \text{ if } L^n_{j,m_1,m_2} = 0,\\
\dfrac{\exp \left(i L^n_{j,m_1,m_2} x_{j+1}\right) - \exp \left(i L^n_{j,m_1,m_2} x_j\right)}{L^n_{j,m_1,m_2}}, & \text{otherwise,}
\end{cases}\\
L^n_{j,m_1,m_2} &= \sum_\beta m_\beta k_\beta \mu_{\beta,j} - nk_0\\
\lambda_{\alpha,j} &= \frac{x_{j + 1}s_{\alpha,j} - x_{j}s_{\alpha, j + 1}}{h_j}, \qquad \mu_{\alpha,j} = \frac{s_{\alpha, j + 1} - s_{\alpha, j}}{h_j}.
\end{split}
\end{equation}
With these approximations in place, the generalized Peierls-Nabarro equation \eqref{eq:pn_hilbert} can be written as
\begin{equation} \label{eq:pn_hilbert_approx_1}
\mathcal{H}\left(\frac{1}{2}\sum_\beta K_{\alpha\beta}\rho_\beta(x)\right) = \sum_{n=-\infty}^{\infty} d_{\alpha,n} \exp (ink_0 x).
\end{equation}
It is emphasized that the foregoing equation is nonlinear on account of the dependence of the coefficients $d_{\alpha,n}$ on the piecewise linear slip distribution $\{s_{\alpha,i}\}$. Using \eqref{eq:ht_sin_cos}, 
the right hand side of \eqref{eq:pn_hilbert_approx_1} can be expressed using Hilbert transforms, thereby resulting in the following equation: 
\begin{equation} \label{eq:pn_hilbert_approx_2}
\mathcal{H}\left(\frac{1}{2}\sum_\beta K_{\alpha\beta}\rho_\beta(x)\right) = \mathcal{H}\left(\sum_n i \, \text{sgn}(n) \, d_{\alpha,n} \exp (ink_0x) \right),
\end{equation}
where $\text{sgn}$ is the sign function. Using \eqref{eq:ht_const} and taking the inverse Hilbert transform of \eqref{eq:pn_hilbert_approx_2} yields the following equation:
\begin{equation} \label{eq:inverse_pn}
\sum_\beta \frac{1}{2}K_{\alpha\beta}\frac{ds_\beta(x)}{dx} = \sum_n i \, \text{sgn}(n) \, d_{\alpha,n} \exp (ink_0x) + a_{\alpha,1},
\end{equation}
where $a_{\alpha,1}$ is a constant that is fixed by the boundary conditions. Since \eqref{eq:inverse_pn} is obtained by discretizing the slip distribution using the piecewise linear approximation \eqref{eq:piecewise_lin_slip}, it is henceforth referred to as the \emph{discrete inverse Peierls-Nabarro model}. A simple algorithm to solve the discrete inverse PN model is presented next.

\subsection{Iterative scheme for the inverse Peierls-Nabarro equation}
The discrete inverse Peierls-Nabarro equation \eqref{eq:inverse_pn} is a nonlinear and non-local equation. One of the advantages of this equation over the original PN equation \eqref{eq:gen_pn} is that integrals with the singular `$1/r$' kernel have been eliminated. A further simplification can be achieved by solving the discrete inverse Peierls-Nabarro equation \eqref{eq:inverse_pn} in an iterative fashion using a relaxed Picard iteration scheme: if $s_\alpha^{(k)}$ denotes the piecewise linear slip distribution at the $k^{\text{th}}$ iteration, then a predictor $\tilde{s}_\alpha^{(k + 1)}$ of the solution of \eqref{eq:inverse_pn} at iteration $k + 1$ can be computed by solving
\begin{equation} \label{eq:ipn_itr}
\sum_\beta \frac{1}{2}K_{\alpha\beta}\frac{d\tilde{s}^{(k + 1)}_\beta(x)}{dx} = \sum_n i \, \text{sgn}(n) \, d_{\alpha,n}(\{s^{(k)}_{\beta,j}\}) \exp (ink_0x) + a_{\alpha,1}.
\end{equation}
Equation \eqref{eq:ipn_itr} can be integrated exactly to yield
\begin{equation} \label{eq:ipn_predictor_1}
\sum_\beta \frac{1}{2} K_{\alpha\beta} \tilde{s}_\beta^{(k + 1)}(x) = \sum_n \frac{\text{sgn}(n)}{nk_0} d_{\alpha,n}(\{s^{(k)}_{\beta,j}\}) \exp (ink_0x) + a_{\alpha,1}x + a_{\alpha,2},
\end{equation}
where $a_{\alpha,2}$ is an arbitrary constant. To fix the constants $a_{\alpha,1}$ and $a_{\alpha,2}$ in the predictor \eqref{eq:ipn_predictor_1}, the following boundary conditions are used
\begin{equation} \label{eq:ipn_bc}
\tilde{s}_{\alpha}^{(k + 1)}(-L) = 0, \qquad \tilde{s}_{\alpha}^{(k + 1)}(L) = S_\alpha.
\end{equation}
Here $S_\alpha$ is the total slip across the slip plane along the $x_\alpha$ direction. Using these boundary conditions, the constants $a_{\alpha,1}$ and $a_{\alpha,2}$ in the predictor \eqref{eq:ipn_predictor_1} are computed as
\begin{equation} \label{eq:ipn_predictor_constants}
\begin{split}
a_{\alpha,1} &= \sum_\beta \frac{1}{2} K_{\alpha\beta} \frac{S_\beta}{2L},\\
b_{\alpha,2} &= \sum_\beta \frac{1}{4} K_{\alpha\beta} S_\beta - \sum_n \frac{(-1)^n\,\text{sgn}(n)}{nk_0} d_{\alpha,n}(\{s^{(k)}_{\beta,j}\}).
\end{split}
\end{equation}
The use of these expressions for the constants in \eqref{eq:ipn_predictor_1} yields the following analytical expression for the the predictor $\tilde{s}^{(k+1)}_\alpha$:
\begin{equation} \label{eq:ipn_predictor_final}
\sum_\beta \frac{1}{2}K_{\alpha\beta} \tilde{s}^{(k + 1)}_\beta(x) = \sum_n \frac{\text{sgn}(n)}{nk_0} \left(d_{\alpha,n}(\{s^{(k)}_{\beta,j}\}) \exp (ink_0x) - (-1)^n\right) + \sum_\beta \frac{1}{2} K_{\alpha\beta} \frac{S_\beta}{2L}(x + L).
\end{equation}
For well understood reasons related to numerical stability, the actual update that yields the slip distribution $s_\alpha^{(k + 1)}$ at iteration $k + 1$ from the slip distribution $s_\alpha^{(k)}$ at iteration $k$ is formulated as
\begin{equation} \label{eq:ipn_update}
s_{\alpha,j}^{(k + 1)} = (1 - \lambda) s_{\alpha,j}^{(k)} + \lambda \tilde{s}_{\alpha,j}^{(k + 1)},
\end{equation} 
where $\lambda \in (0,1]$ is a (small) parameter (see Appendix~\ref{app:anderson_accn_fpi}). The iterations are carried out until $\sum_\alpha \left(\sum_j (s^{k + 1}_{\alpha, j} - s^{k}_{\alpha, j})^2\right)^{\frac{1}{2}} < \text{TOL}$, where TOL is a specified tolerance. It is emphasized that the foregoing iterative strategy is local at every iteration, and is thus very efficient. Notably, this iterative scheme is matrix-free and does not require the formulation of a linear algebraic system of equations, unlike existing techniques to solve the generalized PN model.

In practice, the fixed point iteration scheme developed above converges slowly. A variety of techniques have been developed to accelerate fixed point iterations; these are briefly reviewed in Appendix~\ref{app:anderson_accn_fpi}. The Anderson acceleration scheme with $n_{AA}$ stages is used in this work to accelerate the convergence of the proposed iterative scheme to solve the generalized PN model \eqref{eq:gen_pn}.

\subsection{Effect of applied stresses}
The effect of an external stress on the slip distribution, while straightforward in the generalized PN model, is not trivial within the framework of the inverse PN model proposed here. This is due to the fact that the Hilbert transform of a constant is zero (see \eqref{eq:ht_const}). A numerical strategy that circumvents this problem is proposed herein.

Taking advantage of the fact that the iterative scheme proposed in this work is defined over the domain $[-L,L]$, the effect of an externally applied stress $F_\alpha$ is modeled as
\begin{equation} \label{eq:ext_stress_ipn}
F_\alpha \chi_{[-L,L]}(x),
\end{equation}
where $\chi_{[a,b]}(x)$ is the characteristic function of the interval $[a,b]$ that is equal to $1$ when $x \in [a,b]$ and $0$ otherwise. Using \eqref{eq:ht_inv_characteristic_fn}, the external stress can be modeled as follows:
\begin{equation} \label{eq:ext_stress_ipn_model}
F_\alpha \chi_{[-L,L]}(x) = \mathcal{H}\left(\frac{F_\alpha}{\pi}\log \left(\frac{L - x}{L + x}\right)\right), \quad x \in [-L,L].
\end{equation}
The generalized PN model \eqref{eq:gen_pn} in the presence of an applied stress $F_\alpha$ takes the form
\begin{equation} \label{eq:gen_pn_F}
\sum_\beta \frac{K_{\alpha\beta}}{2\pi}\int_{-\infty}^{\infty} \frac{\rho_\beta(x')}{x - x'} \, dx' = \tau_\alpha(s_1(x), s_2(x)) + F_\alpha.
\end{equation}
Following the same procedure outlined earlier in obtaining \eqref{eq:pn_hilbert_approx_2} from \eqref{eq:gen_pn}, \eqref{eq:gen_pn_F} can be written as
\begin{equation} \label{eq:ipn_F}
\mathcal{H}\left(\frac{1}{2}\sum_\beta K_{\alpha\beta}\rho_\beta(x)\right) = \mathcal{H}\left(\sum_n i \, \text{sgn}(n) \, d_{\alpha,n} \exp (ink_0x)\right) + \mathcal{H}\left(\frac{F_\alpha}{\pi}\log \left(\frac{L - x}{L + x}\right)\right).
\end{equation}
Taking the inverse transform, the resulting inverse PN model can be solved analytically as before at each iteration of the iterative scheme.

While this strategy provides a way to handle external stresses within the current framework, it has its limitations owing to the fact that the domain size $L$ explicitly figures in \eqref{eq:ipn_F}. The specific choice \eqref{eq:ext_stress_ipn} used to model external stresses requires careful handling, especially close to the boundaries. This and related details are discussed in a later section.

\section{Numerical examples}
A few numerical examples are now presented to illustrate the utility of the proposed iterative scheme for the inverse Peierls-Nabarro model. This includes simple test cases using the original PN model \eqref{eq:pn_iso_1d}, dissociated dislocations in one dimension, the equilibration of a dipole to illustrate the handling of external stresses, and a study of the core structure of edge and screw dislocations on the close-packed $\{111\}$ planes in Aluminium.

In all the simulations presented below, $h$ refers to the mesh discretization size, and is defined as $h = \text{max}(h_j)$. When reporting the results, two choices of  discretization, namely a fine discretization $h \to 0$ and a coarse discretization with the atomic scale periodicity $h_j = b$ are chosen. The dislocation density in each case is computed numerically from the slip distribution data by employing a central difference approximation.

\subsection{Original PN model}
\begin{figure}[h!]
\centering
\subfloat[Slip distribution.]{%
\resizebox*{7cm}{!}{\includegraphics{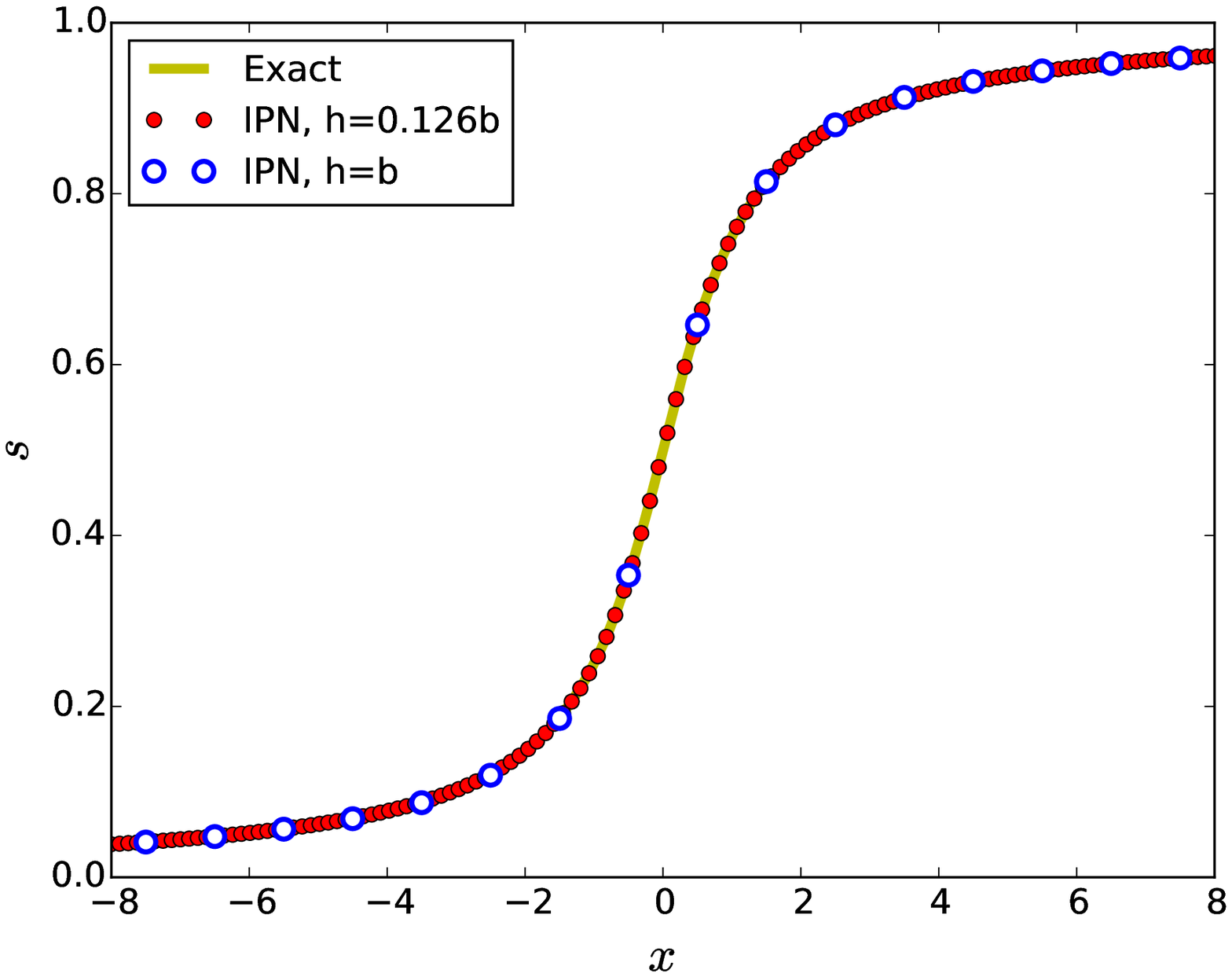}}}\hspace{3pt}
\subfloat[Dislocation density.]{%
\resizebox*{7cm}{!}{\includegraphics{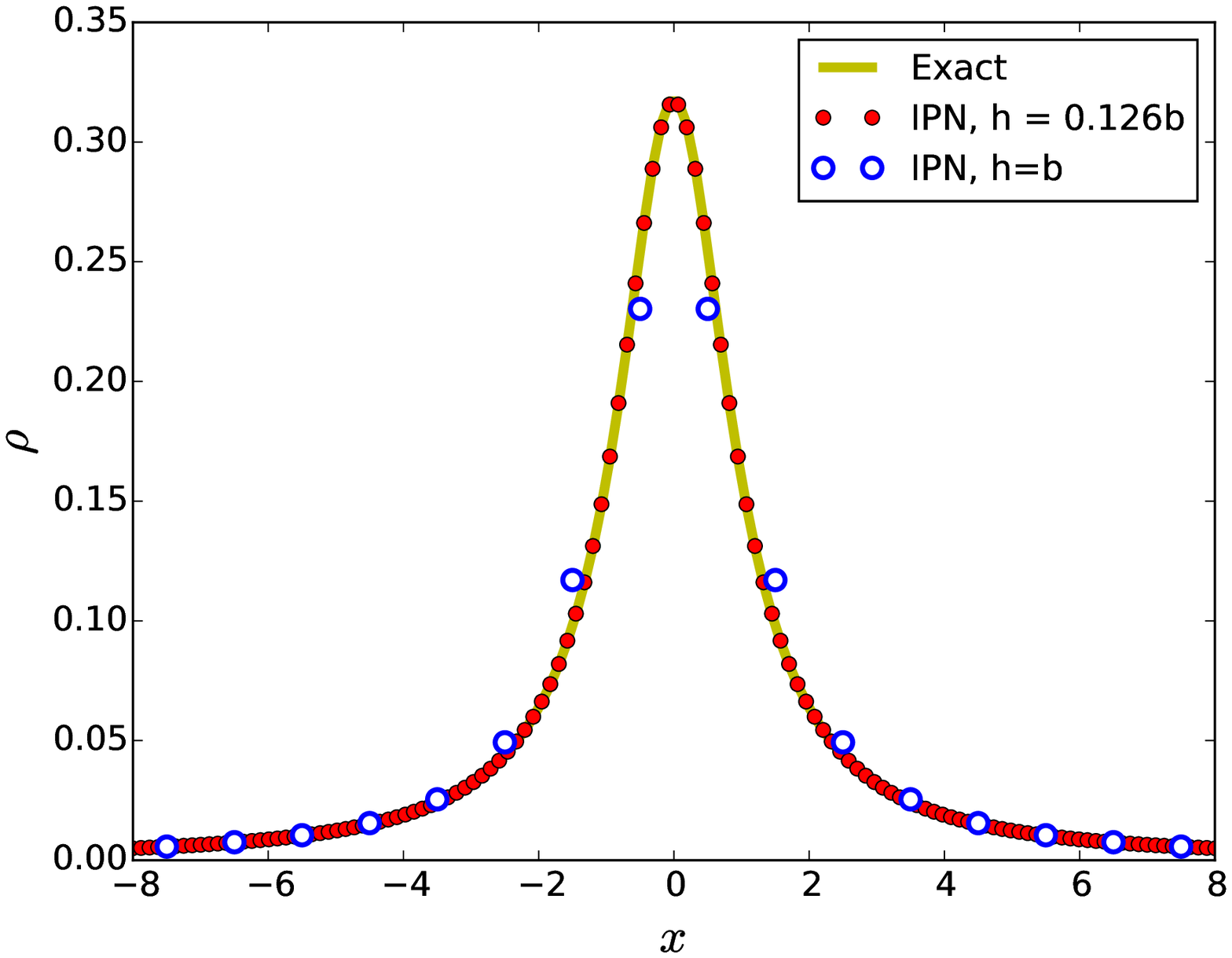}}}
\caption{Comparison of exact solution of the 1d PN model \eqref{eq:pn_iso_1d_soln} and the semi-analytical iterative solution of the inverse PN model. The numerical solutions correspond to a coarse discretization $h = b$ and a fine, but unphysical, discretization $h \simeq b/8$. Here, and in the following, IPN refers to the discrete inverse Peierls-Nabarro model \eqref{eq:inverse_pn}.} 
\label{fig:comp_pn1d_exact}
\end{figure}

The availability of an exact analytical solution for the simple case of a sinusoidal misfit stress, as in \eqref{eq:pn_iso_1d_soln}, provides a simple benchmark to test the proposed numerical strategy to solve the PN equation. A comparison of the slip distribution and the dislocation density obtained by the iterative solution of the inverse PN model and the exact analytical solution is shown in Figure~\ref{fig:comp_pn1d_exact}.
The numerical solution was obtained with the following choice of parameters: $b = 1.0$, $\xi = 1.0$, $L = 50.5$, $N_\tau = 202$ and $\lambda = 0.01$. Two approximate solutions, one corresponding to a fine discretization with $N_s = 801$, and a coarse discretization with $N_s = 101$, were considered. $N_s = 801$ corresponds to a mesh size of approximately $b/8$, while $N_s$ corresponds to the physically relevant mesh size of $b$. The initial guess for the slip distribution was chosen as a step function with slip equal to $0$ in $[-L,0)$, and equal to $b$ in $[0,L]$. The iterations were carried out until the error in the slip update is less than a specified tolerance level of $\text{TOL} = 10^{-6}$. It is seen that both the fine and coarse meshes produce results in good agreement with the exact solution, with the finer mesh being more accurate than the coarser mesh - this will be discussed in more detail shortly.

\begin{figure}[h!]
\centering
\includegraphics[width=0.6\textwidth]{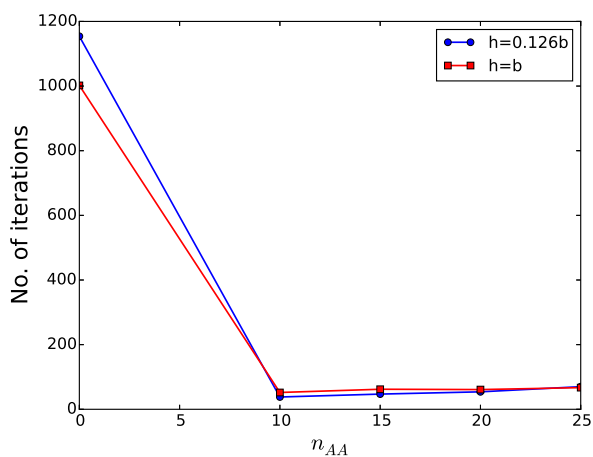}
\caption{Number of iterations for convergence of the Anderson acceleration scheme for fixed point iteration of the inverse PN model with $n_{AA}$-stages for both the fine and coarse mesh discretizations.}
\label{fig:conv_ipn1d_AA}
\end{figure}

The naive fixed point iteration used in obtaining Figure~\ref{fig:comp_pn1d_exact} requied $1002$ iterations when $h=b$, $1154$ iterations when $h\simeq b/8$. To accelerate the convergence, the Anderson acceleration technique for fixed point iteration (see Appendix~\ref{app:anderson_accn_fpi}) with $n_{AA}$ stages was used. The number of iterations required to reach a tolerance level $\text{TOL} < 10^{-6}$ is shown in Table~\ref{tab:conv_ipn1d_AA} and Figure~\ref{fig:conv_ipn1d_AA}. The other parameters were kept fixed at the same values as before: $b = 1.0$, $\xi = 1.0$, $L = 50.5$, $N_\tau = 202$, and $N_s = 101, 801$. The starting guess for the slip distribution was set as the same step distribution as before. It is seen that there is an optimal value of $n_{AA}$, in this case close to $10$, for which the number of iterations for convergence is minimum. This value, however, is dependent on the problem at hand. The key observation is that the number of iterations with Anderson acceleration is $O(10)$ - a significant improvement compared to the $O(10^3)$ iterations required for convergence in the naive iteration scheme. For this reason, all numerical results presented henceforth are those obtained using the Anderson accelerated fixed point iterations.

\begin{table}[h!]
\tbl{Convergence of Anderson acceleration scheme. $n_{AA}=0$ corresponds to the naive fixed point iteration.}
{\begin{tabular}{clllll} \toprule
$n_{AA}$ & 0 & 10 & 15 & 20 & 25\\ \midrule
$h=b$ & 1154 & 38 & 47 & 54 & 70\\
$h\simeq b/8$ & 1002 & 52 & 62 & 61 & 67\\ \bottomrule
\end{tabular}}
\label{tab:conv_ipn1d_AA}
\end{table}

\begin{table}
\tbl{Error in the numerical solution of the 1D inverse PN model for sinusoidal misfit stresses as a function of domain size.}
{\begin{tabular}{lllll} \toprule
 & \multicolumn{2}{c}{$h = b$} & \multicolumn{2}{c}{$h \simeq b/8$}\\ \cmidrule{2-3} \cmidrule{4-5}
$L$ & $\epsilon_{10b}$ & Iterations & $\epsilon_{10b}$ & Iterations\\ \midrule
10.5 & 1.587 $\times 10^{-2}$ & 38 & 1.569 $\times 10^{-2}$ & 34\\ 
20.5 & 4.204 $\times 10^{-3}$ & 36 & 3.849 $\times 10^{-3}$ & 31\\
30.5 & 2.136 $\times 10^{-3}$ & 45 & 1.733 $\times 10^{-3}$ & 36\\
40.5 & 1.451 $\times 10^{-3}$ & 46 & 9.776 $\times 10^{-4}$ & 38\\
50.5 & 1.160 $\times 10^{-3}$ & 52 & 6.280 $\times 10^{-4}$ & 38\\
60.5 & 1.016 $\times 10^{-3}$ & 68 & 4.386 $\times 10^{-4}$ & 54\\
80.5 & 8.909 $\times 10^{-4}$ & 116 & 2.530 $\times 10^{-4}$ & 48\\
100.5 & 8.407 $\times 10^{-4}$ & 52 & 1.640 $\times 10^{-4}$ & 106\\ \bottomrule
\end{tabular}}
\label{tab:conv_ipn1d_L}
\end{table}

\begin{figure}[h!]
\centering
\includegraphics[width=0.6\textwidth]{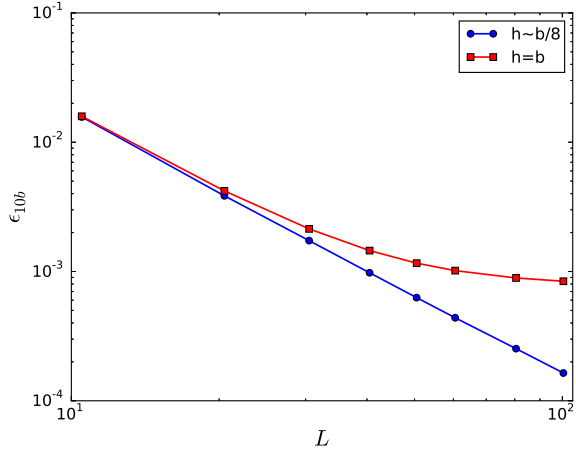}
\caption{Error in the interval $[-10b,10b]$ between exact and iterative solutions of \eqref{eq:pn_iso_1d} as a function of domain size $L$ (in units of $b$). Errors for both the fine discretization ($h \simeq b/8$) and the coarse discretization ($h = b$) are shown.}
\label{fig:conv_ipn1d_L}
\end{figure}

To monitor the effect of the domain size $L$ on the accuracy of the iterative solution, the error $\epsilon$ between the iterative and exact solutions, defined as
\begin{displaymath}
\epsilon_{\tilde{L}}^2 = \frac{1}{N_{I_{\tilde{L}}}b^2}\sum_{i \in I_{\tilde{L}}} (s_i - s_\text{exact}(x_i))^2,
\end{displaymath}
is plotted as a function of $L$ in Figure~\ref{fig:conv_ipn1d_L}, and listed in Table~\ref{tab:conv_ipn1d_L}, for both the fine ($h \simeq b/8$) and coarse ($h = b$) mesh discretizations. Here $I_{\tilde{L}}$ is the set of indices with the property that $x_i \in [-\tilde{L}, \tilde{L}]$ whenever $i \in I$, and $N_{I_{\tilde{L}}}$ is the number of such indices. For concreteness, $\tilde{L} = 10b$ was chosen in computing the errors shown in Figure~\ref{fig:conv_ipn1d_L}. The various constants are chosen as follows: $b = 1.0$, $\xi = 1.0$, $N_\tau = 2(2L/b)$, and $n_{AA} = 10$. $N_s = 2L/b$ for the coarse mesh, and $N_s = (16L/b) - 7$ for the fine mesh. The number of iterations required for convergence is also listed in  Table~\ref{tab:conv_ipn1d_L}. As can be seen from Figure~\ref{fig:conv_ipn1d_L} and Table~\ref{tab:conv_ipn1d_L}, the error associated with the fine discretization decreases rapidly with the domain size, while the error in the coarse mesh discretization initially decreases rapidly, but eventually decreases slowly with increase in domain size.
\begin{figure}[h!]
\centering
\includegraphics[width=0.6\textwidth]{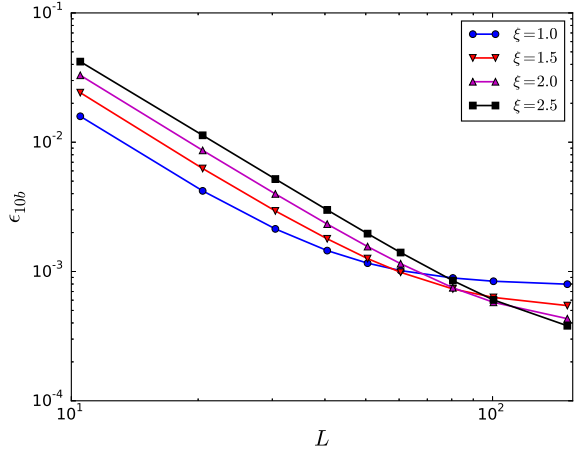}
\caption{Error in $[-10b,10b]$ between exact and iterative solutions of \eqref{eq:pn_iso_1d} as a function of domain size $L$ (in units of $b$) for various values of the dislocation profile width $\xi$. The coarse discretization ($h = b$) was used for all these simulations.}
\label{fig:conv_ipn1d_xi}
\end{figure}
To understand the origin of the slower decrease in the error associated with the coarse discretization $h = b$, as seen in Figure~\ref{fig:conv_ipn1d_L}, a plot of the error $\epsilon_{10b}$ is plotted as a function of the domain size $L$ for various values of the dislocation width $\xi$ in Figure~\ref{fig:conv_ipn1d_xi}. The various parameters used in obtaining Figure~\ref{fig:conv_ipn1d_xi} are the same as those used for the simulations in Figure~\ref{fig:conv_ipn1d_L}. It can be seen that the error corresponding to wider dislocation profiles is lower than that of narrower dislocation profiles when the domain size is sufficiently large. Stated differently, the error associated with the coarse mesh discretization is lower when $b/\xi$ is small, or, equivalently, when the slip distribution varies slowly at the scale of the atomic lattice.

\subsection{Dissociated dislocations in 1d}
\begin{figure}
\centering
\subfloat[Slip distribution.]{%
\resizebox*{7cm}{!}{\includegraphics{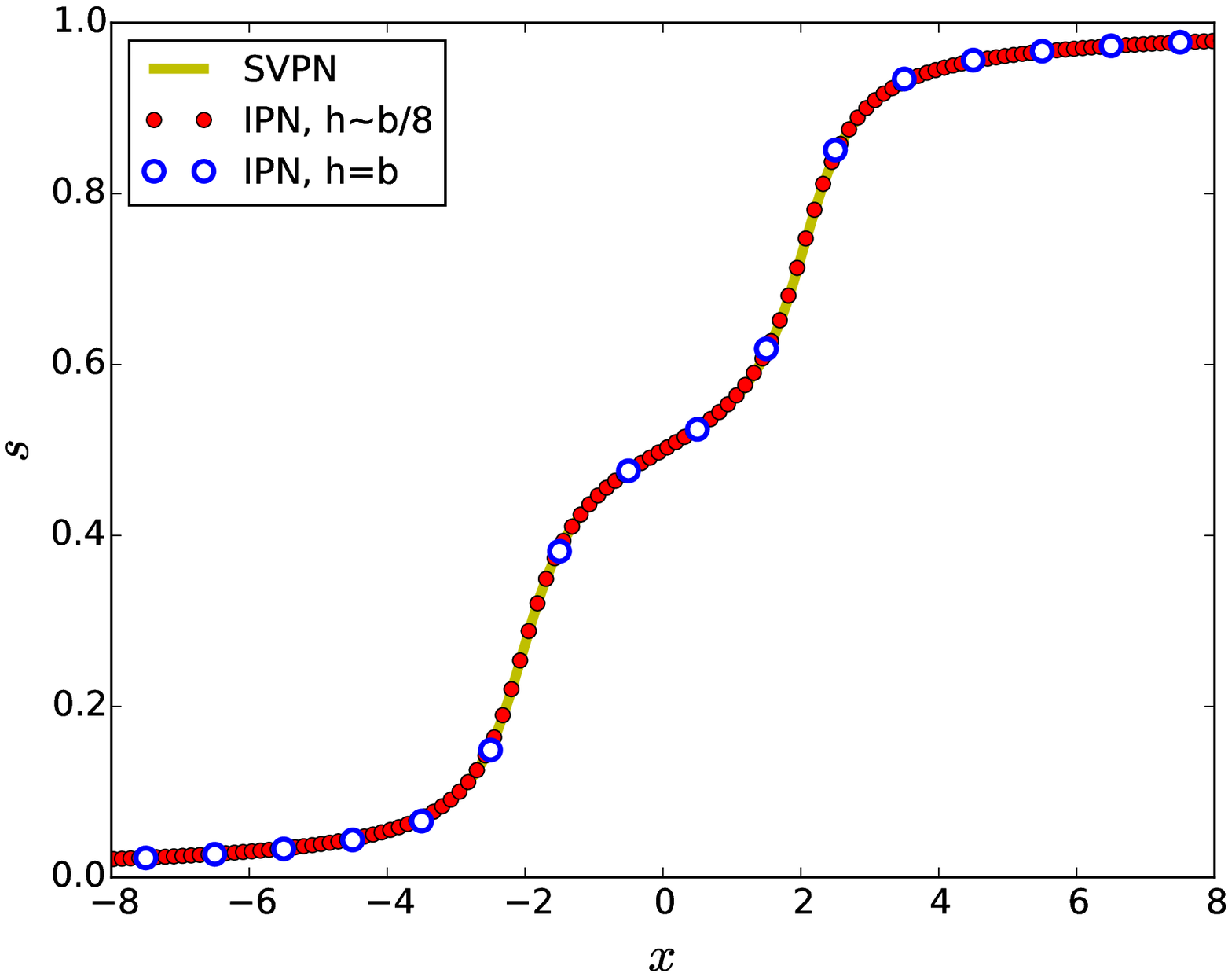}}}\hspace{3pt}
\subfloat[Dislocation density.]{%
\resizebox*{7cm}{!}{\includegraphics{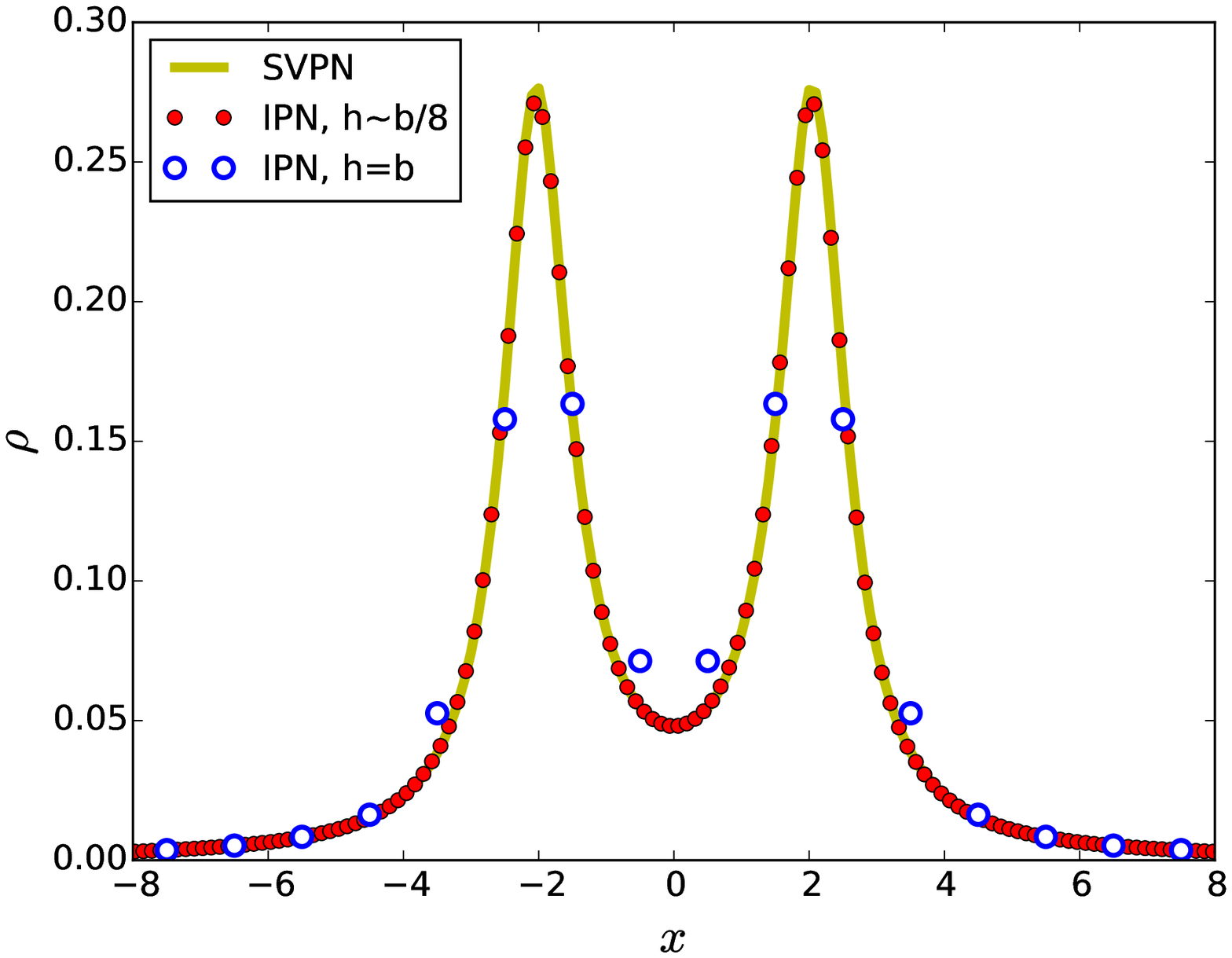}}}
\caption{Comparison of IPN and SVPN for dissociated dislocation core. The iterative solution corresponding to both coarse ($h = b$) and fine ($h \simeq b/8$) mesh discretizations are shown.} 
\label{fig:comp_pn1d_partials}
\end{figure}

The slip distribution corresponding to a 1d PN model with a more complex slip distribution is considered in this section. Specifically, the misfit stress is chosen to be of the form
\begin{displaymath}
\tau(\delta) = \frac{\mu}{2\pi d}\left(\sin \frac{2\pi \delta}{b} + 2\sin \frac{4\pi \delta}{b}\right).
\end{displaymath}
In this case, there is no exact analytical solution available. Further, the presence of a local minimum in the misfit stress distribution is expected to cause a dissociation of the dislocation into partial dislocations separated by a finite width stacking fault. To validate the solution obtained by the proposed method, the 1D PN model was also solved using the semi-discrete variational Peierls-Nabarro (SVPN) model \cite{BK97}. A comparison of the solution obtained by these two methods is shown in Fig.~\ref{fig:comp_pn1d_partials}.
The following parameters were used for the reference SVPN solution: $b = 1.0$, $\xi = 2.5$, $L = 50.0$, $N_s = 2000$. For the inverse PN model, the following parameters were used: $L = 50.5$, $N_s = 201, 1601$, $N_\tau = 402$, $\lambda = 0.01$ and $n_{AA} = 20$. It can be seen that the proposed method agrees well with the dissociated core structure that is computed using the SVPN method. This example is included to highlight the fact that the proposed iterative scheme for the inverse PN model is competitive with existing techniques like the SVPN method for non-trivial misfit stresses.

\subsection{Effect of external stress}
As mentioned earlier, the fact that the Hilbert transform removes the \emph{DC} component of the function it acts on requires that external stresses are handled in a non-standard manner. Specifically, the effect of an external stress is represented by a term of the form $F \chi_{[-L,L]}$. However, it is to be noted that since the current formulation is based on the original PN model, it inherits one its fundamental limitations: the energy of an infinite linear elastic medium with a single dislocation is independent of the position of the dislocation. The fact that the present formulation enforces the slip to have definite values at the boundaries of a finite domain $[-L,L]$ introduces, however, an artificial dependence of the energy on the slip distribution over the finite domain $[-L,L]$ via a confining potential, whose origin lies in the choice of the approximation \eqref{eq:ext_stress_ipn} for external stresses within the current framework.
\begin{figure}[h!]
\centering
\includegraphics[width=0.6\textwidth]{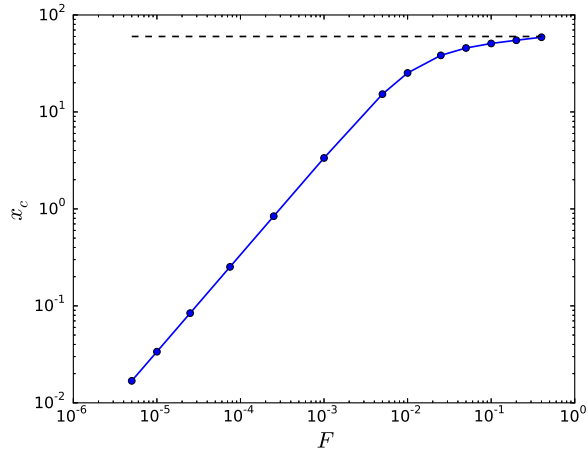}
\caption{Dependence of the center of the dislocation \eqref{eq:disl_center} on the applied force. The position of the domain boundary is indicated as a dashed black line.}
\label{fig:pn1d_force_displ}
\end{figure}
This is illustrated in Figure~\ref{fig:pn1d_force_displ}, where the center of the dislocation, defined as
\begin{equation} \label{eq:disl_center}
x_c = \int_{-L}^L x \rho(x) \, dx,
\end{equation}
is plotted as a function of the external force $F$. The graph shown in Figure~\ref{fig:pn1d_force_displ} was obtained by solving \eqref{eq:ipn_F} with sinusoidal misfit stress, and with the following set of parameters: $b = 1$, $\xi = 1.0$, $L = 60.0$, $N_s = 1201$, $N_\tau = 240$, and $n_{AA} = 20$. The slope of the initial part of the force-displacement curve in Figure~\ref{fig:pn1d_force_displ} is $1.0$ (within numerical errors). Thus, the dislocation experiences no lattice resistance, as expected in the original PN model. The slope decreases towards zero as the dislocation approaches the boundary, since the boundary conditions require the slip to be zero on the boundaries, and hence prevent further dislocation motion. The non-linearity in the force-displacement curve in Figure~\ref{fig:pn1d_force_displ} is thus a direct consequence of the finiteness of the domain considered. The fact that the dislocation has a definite position for a given valued of the external stress illustrates the artificial confining potential introduced by modeling the external stress as in \eqref{eq:ext_stress_ipn}. The foregoing result also shows that the Peierls stress cannot be calculated within the present approach. The Peierls stress may be calculated as in the conventional procedure by using the discrete solution obtained by the present formulation in a discrete approximation of the energy which recovers the effect of lattice periodicity \cite{nabarro52}. The computation of the Peierls stress along these lines is not carried out in this work. 

To illustrate the foregoing comments on handling external stresses within the current formulation, as in \eqref{eq:ipn_F}, the equilibration of a dislocation dipole is considered. Consider two edge dislocations of equal and opposite Burgers vector $b$ separated by a distance $2w$ on a glide plane. In the absence of an external stress, the linear elastic theory predicts that this dipole will annihilate itself. The stress $\sigma_e$ required to stabilize this dipole can be easily computed using the linear elastic theory as
\begin{equation} \label{eq:stress_dipole}
\sigma_e = \frac{\mu\xi}{2\pi d w}.
\end{equation}
\begin{table}
\tbl{Convergence of equilibration stress, in units of $\pi d/\mu\xi$, with respect to domain size $L$ for a dipole of width $2w = 80.0$ computed using the IPN model.}
{\begin{tabular}{lp{3cm}} \toprule
$L$ & $\sigma_e$\\ \midrule
75.0 & 1.75 $\times 10^{-2}$\\
100.0 & 2.15 $\times 10^{-2}$\\
150.0 & 2.53 $\times 10^{-2}$\\
200.0 & 2.68 $\times 10^{-2}$\\
250.0 & 2.70 $\times 10^{-2}$\\ \bottomrule
\end{tabular}}
\label{tab:pn1d_dipole_conv}
\end{table}
\begin{figure}[h!]
\centering
\includegraphics[width=0.6\textwidth]{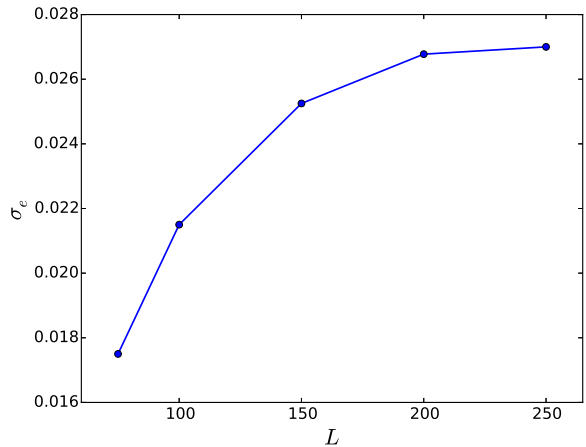}
\caption{Convergence with respect to domain size of the (scaled) external stress for equilibration of a dipole.}
\label{fig:pn1d_dipole_conv}
\end{figure}
The equilibration of the dipole is now reconsidered using the inverse PN model \eqref{eq:ipn_F} using the sinusoidal stacking fault energy \eqref{eq:gamma_sin_1d}. Fixing the width of the dipole to $w = 40.0$, the dependence of the equilibration stress as a function of the domain size is shown in Figure~\ref{fig:pn1d_dipole_conv} and Table~\ref{tab:pn1d_dipole_conv}. The following parameters were chosen: $N_s = 10(2L/b) + 1$, $N_\tau = 4L/b$, $n_{AA} = 10$ and $\lambda = 0.01$. It is seen that the stress required for equilibration converges if the domain is large enough.

\begin{table}
\tbl{Equilibration stress, in units of $\pi d/\mu \xi$, for a dipole of width $2w$ computed using both linear elasticity and the IPN model.}
{\begin{tabular}{clllllll} \toprule
& \multicolumn{6}{c}{$w$} \\ \cmidrule{2-8}
& 5.0 & 10.0 & 20.0 & 40.0 & 60.0 & 80.0 & 100.0 \\ \midrule
Linear Elasticity & 2 $\times 10^{-1}$ & 1 $\times 10^{-1}$ & 5 $\times 10^{-2}$ & 2.5 $\times 10^{-2}$ & 1.67 $\times 10^{-2}$ & 1.25 $\times 10^{-2}$ & 1 $\times 10^{-2}$\\
Inverse PN model & 2.5 $\times 10^{-1}$ & 1.2 $\times 10^{-1}$ & 5.75 $\times 10^{-2}$ & 2.68 $\times 10^{-2}$ & 1.57 $\times 10^{-2}$ & 9.63 $\times 10^{-3}$ & 5.5 $\times 10^{-3}$\\ \bottomrule
\end{tabular}}
\label{tab:dipole_stress_eqbm}
\end{table}
\begin{figure}[h!]
\centering
\includegraphics[width=0.6\textwidth]{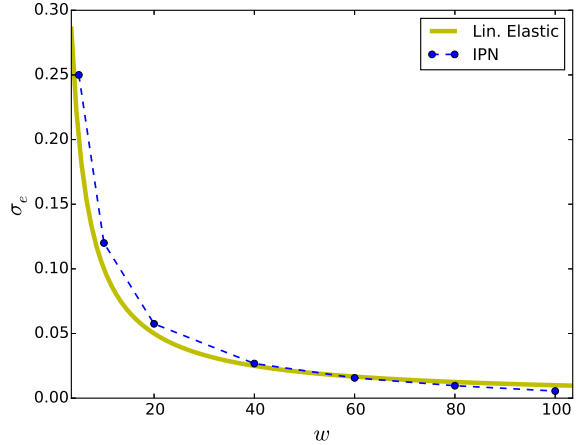}
\caption{Comparison of the (scaled) external stress required for equilibration of dipole computed according to linear elasticity and the Peierls-Nabarro model with sinusoidal misfit stresses.}
\label{fig:pn1d_dipole_stress}
\end{figure}
A comparison of the (scaled) external stress required for equilibration of the dipole as a function of its width is shown in Figure~\ref{fig:pn1d_dipole_stress} and Table~\ref{tab:dipole_stress_eqbm}. The following parameters were used: $b = 1$, $\xi = 1$, $L = 200$, $N_s = 4001$, $N_t = 800$, $\lambda = 0.01$ and $n_{AA} = 20$. It can be seen that the predictions of the sinusoidal PN model, computed using the iterative scheme proposed in this work, are in qualitative agreement with the predictions of linear elasticity. The expected deviations from the linear elastic theory that is seen in Figure~\ref{fig:pn1d_dipole_stress} arise due to two distinct sources. First, there is a modeling error due to the fact that the PN solution is computed assuming that the stacking fault energy is sinusoidal in nature, as in \eqref{eq:gamma_sin_1d}. Second, referring to Figure~\ref{fig:pn1d_dipole_conv}, the error associated with smaller dipole widths are smaller since $L/w$ is larger than that for dipoles with larger width. As a final comment, it is important to place the dipole at the center of the domain since boundary effects that arise due to the specific manner in which external stresses are modeled here become important as the dipole approaches the boundaries.

\subsection{Core structure of edge and screw dislocations in Aluminium}
While the foregoing examples involved toy models with one-component slip distributions, a more realistic problem concerning the core structure of both edge and screw dislocations in Aluminium is now analyzed using the inverse Peierls-Nabarro model. The core structure of dislocations in Al is well studied (see, for instance, \cite{schoeck01}, \cite{schoeck02}, \cite{schoeck12}, \cite{RVCPW17}, \cite{LCWCS17}), and hence serves as a good benchmark to validate the proposed iterative method to solve the generalized PN model. As is well known, perfect dislocations on the $\{111\}$ planes in FCC materials split into Shockley partials for energetic reasons \cite{AHL16}. One of the goals of this study is to reproduce the core structure of these dissociated dislocations and compare them with results obtained earlier.

\begin{figure}[h!]
\centering
\includegraphics[width=0.7\textwidth]{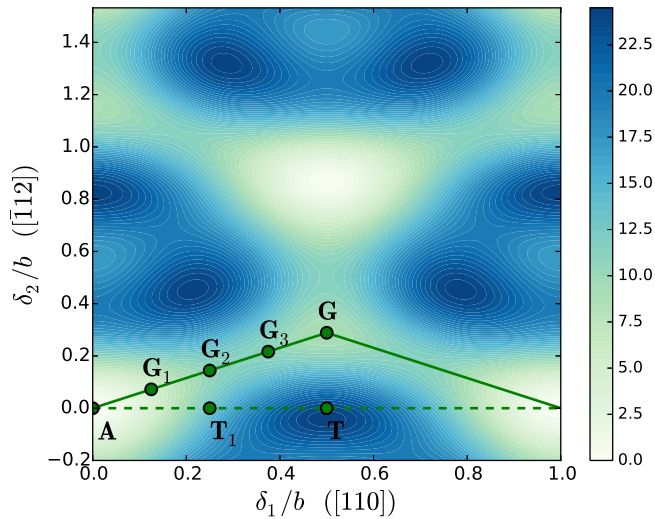}
\caption{Contour plot of stacking fault energy (in meV/\AA${}^2$) of $\{111\}$ planes in Aluminium. The plot was generated using \eqref{eq:gamma_Al} and the constants listed in Table~\ref{tab:gamma_Al_constants}. The data points listed in Table~\ref{tab:gsfe_Al_data} is also shown for reference. The dashed line indicates the direction of the perfect dislocation, while the thick lines indicate the directions along the Schockley partials into which the perfect dislocation dissociates.}
\label{fig:gsfe_Al_111}
\end{figure}

For concreteness, dislocations on the close-packed $\{111\}$ planes in Aluminium were considered. Specifically, edge dislocations, with line direction $t = [\bar{1}12]$, and screw dislocations, with line direction $t = [110]$, were studied. The symmetric anisotropic Stroh tensor \cite{BBS79} is used to model the elastic interactions. If $n$ denotes the normal to the $\{111\}$ plane on which the dislocation resides, the direction $m$ perpendicular to the dislocation line is given by the vector cross product of $n$ and $t$: $m = n \times t$. The components of the symmetric anisotropic Stroh tensor on the $\{111\}$ plane for Aluminium are tabulated in Table~\ref{tab:stroh_Al}; these values are based on those reported in \cite{SHLB18}. 
\begin{table}
\tbl{Components of the symmetric anisotropic Stroh tensor on $\{111\}$ slip planes for edge and screw dislocations in Al \cite{SHLB18}. All values are in meV/\AA${}^3$. It is to be noted that the line directions of the edge and screw dislocations are different, as indicated, but they both lie in the same $\{111\}$ slip plane.}
{\begin{tabular}{clll} \toprule
Type & $K_{mm}$ & $K_{tt}$ & $K_{tm} (= K_{mt})$ \\ \midrule
Edge, $t = [\bar{1}12]$ & 261.444 & 167.247 & 0.0\\
Screw, $t = [110]$ & 261.444 & 167.247 & 0.0\\ \bottomrule
\end{tabular}}
\label{tab:stroh_Al}
\end{table}
The generalized stacking fault energy surface is computed using the fitting procedure outlined in \cite{schoeck01}. The generalized stacking fault energy on $\{111\}$ planes is taken to be of the form
\begin{equation} \label{eq:gamma_Al}
\begin{split}
\gamma(\delta_1, \delta_2) &= c_0 + 2c_1 \cos k_1\delta_1 \cos k_2\delta_2 + c_1 \cos 2k_2\delta_2 + c_2 \cos 2k_1\delta_1\\
 & + 2c_2 \cos k_1\delta_1 \cos 3k_2\delta_2 + c_3 \cos 4k_2\delta_2 + 2c_3 \cos 2k_1\delta_1 \cos 2k_2\delta_2\\
 & + 2c_4 \cos 3k_1\delta_1 \cos k_2\delta_2 + 2c_4 \cos 2k_1\delta_1 \cos 4k_2\delta_2 + 2c_4 \cos k_1\delta_1 \cos 5k_2\delta_2\\
 & + a_1 \sin 2k_2\delta_2 + a_3 \sin 4k_2\delta_2 - 2a_1 \cos k_1\delta_1 \sin k_2\delta_2 - 2a_3 \cos 2k_1\delta_1 \sin 2k_2\delta_2, 
\end{split}
\end{equation}
where
\begin{equation} \label{eq:def_k1k2}
k_1 = \frac{2\pi}{b}, \qquad k_2 = \frac{2\pi}{\sqrt{3}b},
\end{equation}
$b = a/\sqrt{2}$, and $a = 4.05$\AA{} is the lattice constant of Al. The various constants in \eqref{eq:gamma_Al} are listed\footnote{Some of the constants are listed incorrectly in \cite{schoeck01}. The corrected version of these constants are reported here.} in Table~\ref{tab:gamma_Al_constants}.
\begin{table}
\tbl{Constants (in meV/\AA${}^2$) for the Fourier approximation, \eqref{eq:gamma_Al}, of the stacking fault energy of $\{111\}$ planes for Aluminium. These constants were computed based on the data given in \cite{schoeck01}. Some of the constants differ from those given in \cite{schoeck01}: see text for details.}
{\begin{tabular}{lllllll} \toprule
$c_0$ & $c_1$ & $c_2$ & $c_3$ & $c_4$ & $a_1$ & $a_3$ \\ \midrule
15.0683 & -3.4607 & -2.1398 & 0.9067 & -0.1645 & -2.1155 & -0.5574\\ \bottomrule
\end{tabular}}
\label{tab:gamma_Al_constants}
\end{table}
These constants are obtained by requiring that the expression \eqref{eq:gamma_Al} for the stacking fault energy agree with known values of the stacking fault energy (computed, for instance, from lower scale models) for specific stacking configurations. These configurations, referred to as $A, T, T_1, G, G_1, G_2, G_3$ following \cite{schoeck01}, are shown in the contour plot of the stacking fault energy on $\{111\}$ planes in Al is shown in Figure~\ref{fig:gsfe_Al_111}. The stacking configurations used to fit the constants in \eqref{eq:gamma_Al} are listed in Table~\ref{tab:gsfe_Al_data} - these values are taken from \cite{schoeck01}. 
\begin{table}
\tbl{Data used to calibrate the constants in \eqref{eq:gamma_Al}. The value for the stacking fault energy corresponding to a specific stacking configuration is based on \cite{schoeck01}.}
{\begin{tabular}{llll} \toprule
Point & $\delta_1/b$ & $\delta_2/b$ & $\gamma(\delta_1, \delta_2)$ (meV/\AA${}^2$)\\ \midrule
$A$ & 0 & 0 & 0.0\\
$G$ & 1/2 & 1/2$\sqrt{3}$ & 8.925\\
$T$ & 1/2 & 0 & 23.718\\
$T_1$ & 1/4 & 0 & 13.170\\
$G_1$ & 1/8 & 1/8$\sqrt{3}$ & 5.056\\
$G_2$ & 1/4 & 1/4$\sqrt{3}$ & 10.673\\
$G_3$ & 3/8 & $\sqrt{3}$/8 & 10.923\\ \bottomrule
\end{tabular}}
\label{tab:gsfe_Al_data}
\end{table}

\begin{figure}
\centering
\subfloat[Slip distribution.]{%
\resizebox*{7cm}{!}{\includegraphics{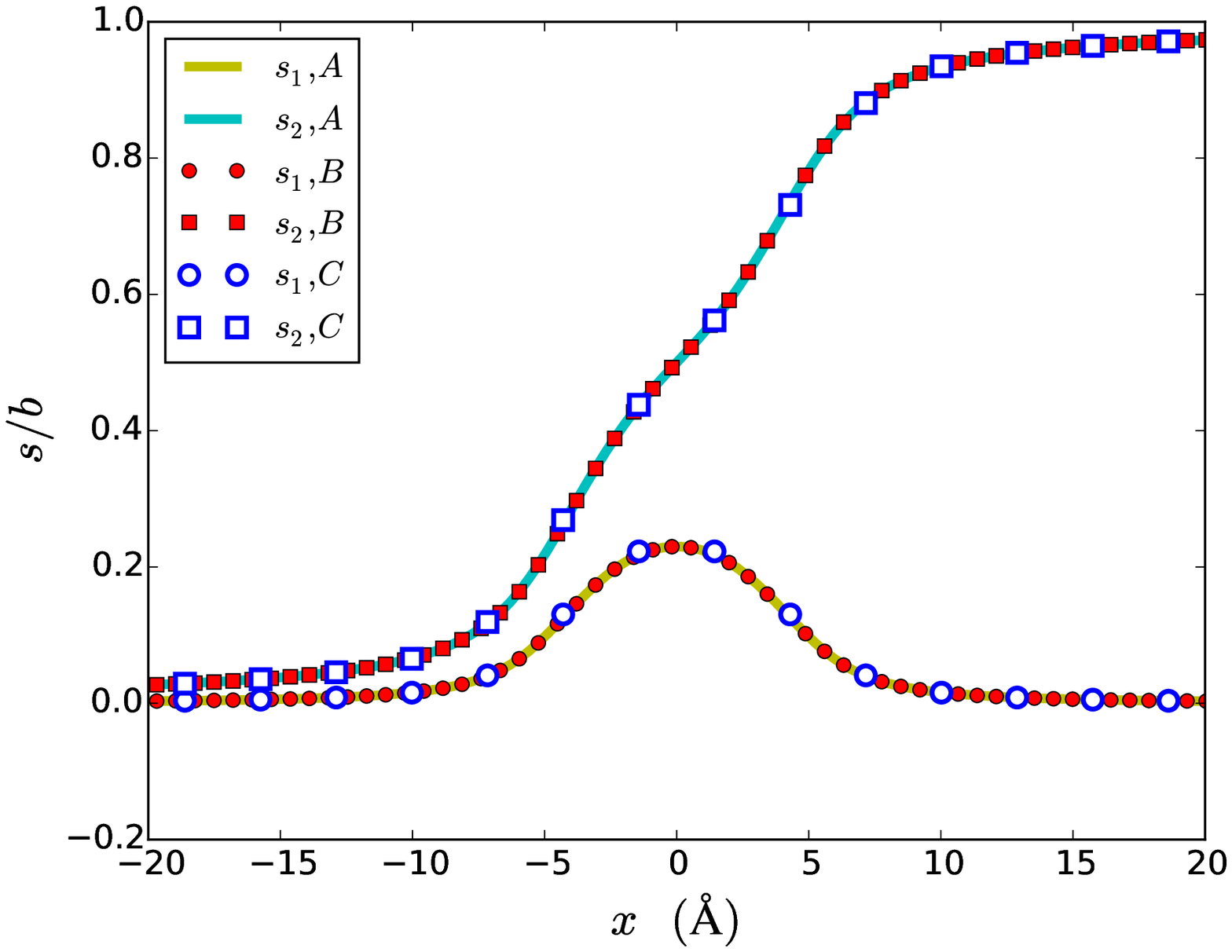}}}\hspace{3pt}
\subfloat[Dislocation density.]{%
\resizebox*{7cm}{!}{\includegraphics{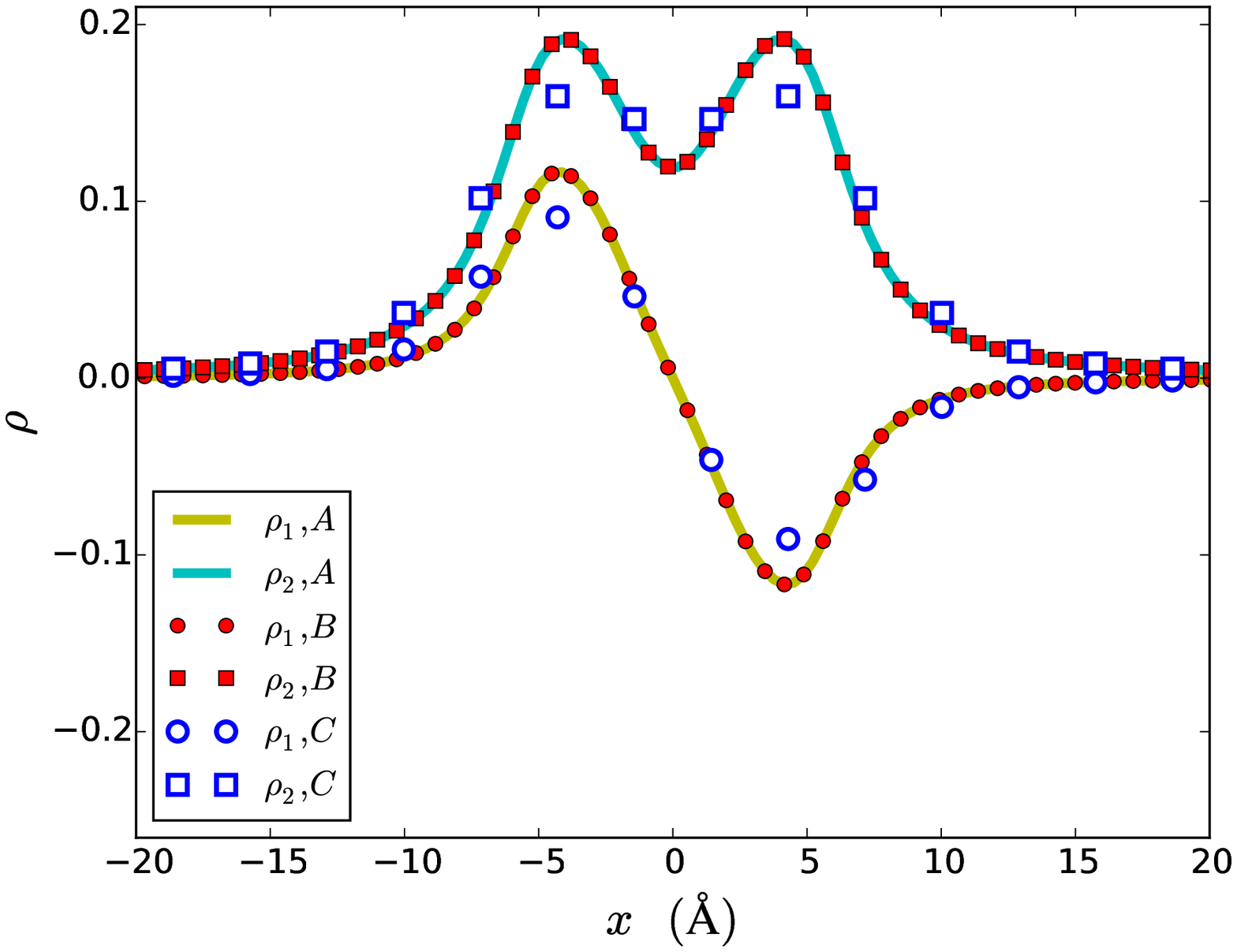}}}
\caption{In-plane components of slip distribution and dislocation density for an edge dislocation on the $(1\bar{1}1)$ plane with line direction $[\bar{1}12]$. The numerical results corresponding to three different discretization schemes: $A: L=75.5b, h \simeq b/8, B: L=50.5b, h \simeq b/8, C: L=50.5b, h=b$ are shown. Here, $s_1$ is the slip along the direction $[110]$ and $s_2$ is the slip along the direction $[\bar{1}12]$. A similar interpretation applies to $\rho_1, \rho_2$. The dislocation density plots were obtained by using a second order finite difference approximation of the corresponding slip distribution data.} 
\label{fig:ipn2d_Al_edge}
\end{figure}

\begin{figure}
\centering
\subfloat[Slip distribution.]{%
\resizebox*{7cm}{!}{\includegraphics{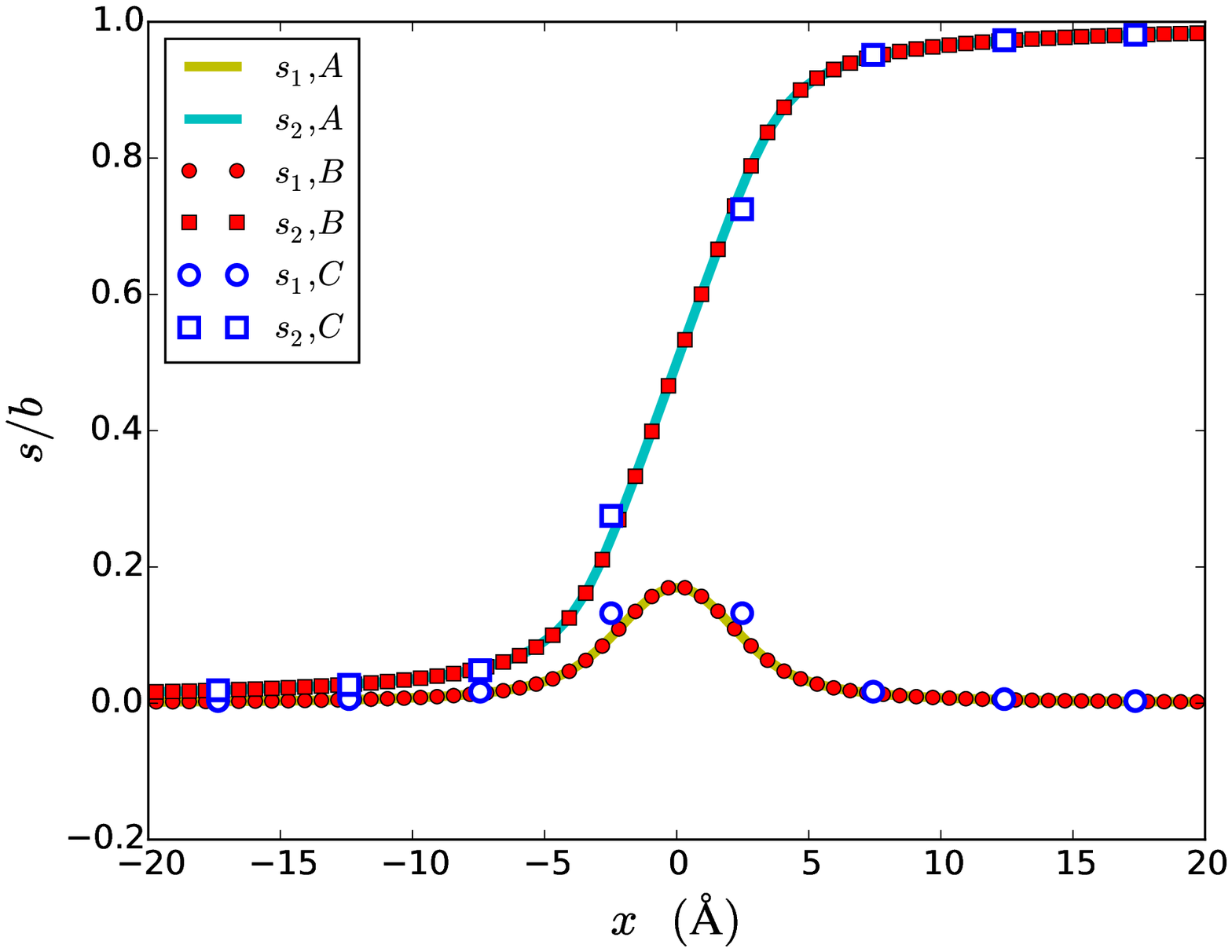}}}\hspace{3pt}
\subfloat[Dislocation density.]{%
\resizebox*{7cm}{!}{\includegraphics{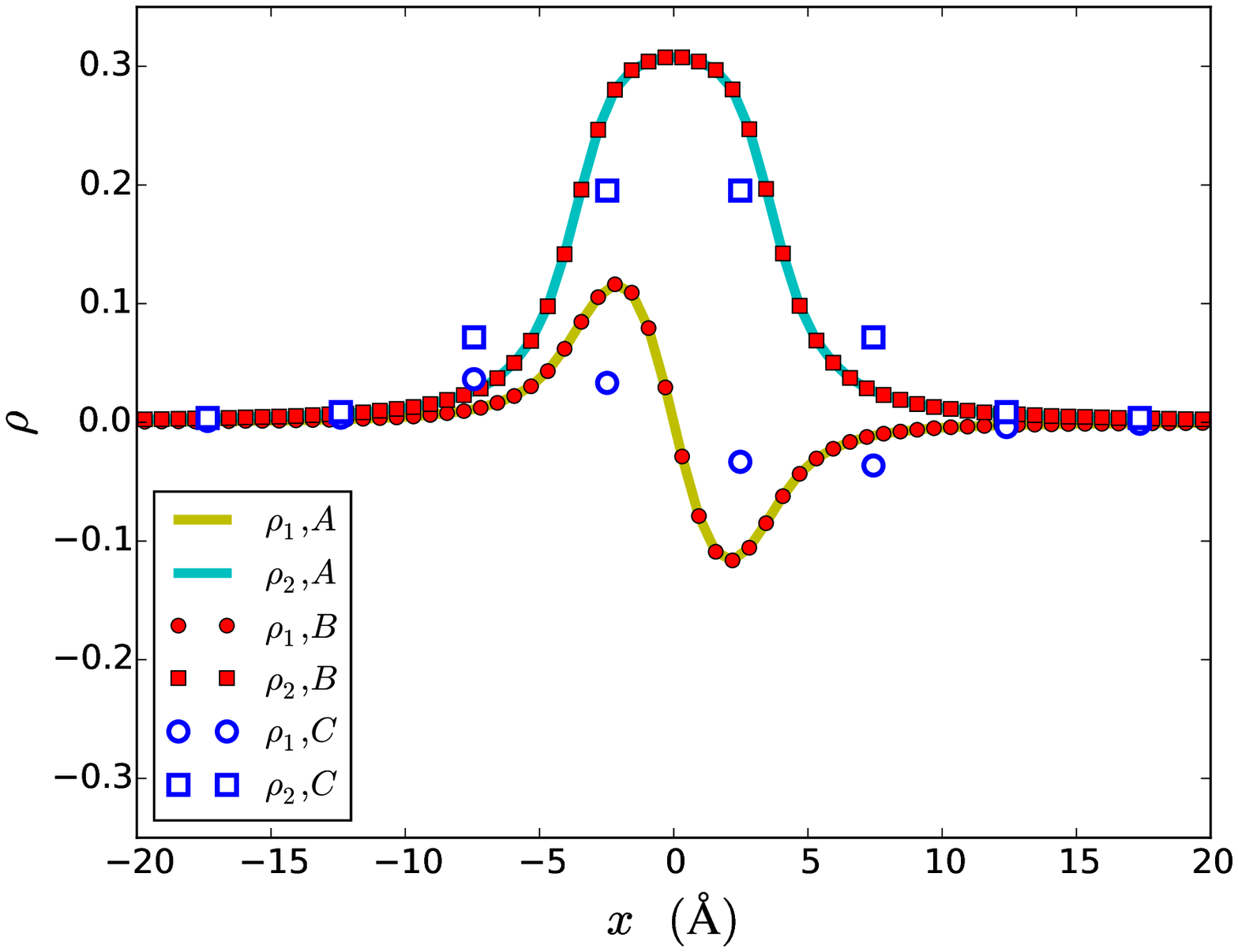}}}
\caption{In-plane components of slip distribution and dislocation density for a screw dislocation on the $(1\bar{1}1)$ plane with line direction $[110]$. The numerical results corresponding to three different discretization schemes: $A: L=75.5\sqrt{3}b, h \simeq \sqrt{3}b/8, B: L=50.5\sqrt{3}b, h \simeq \sqrt{3}b/8, C: L=50.5\sqrt{3}b, h = \sqrt{3}b$ are shown. Here, $s_1$ is the slip along the direction $[\bar{1}12]$ and $s_2$ is the slip along the direction $[110]$. A similar interpretation applies to $\rho_1, \rho_2$. The dislocation density plots were obtained by using a second order finite difference approximation of the corresponding slip distribution data.} 
\label{fig:ipn2d_Al_screw}
\end{figure}

The slip and dislocation density distributions corresponding to an edge dislocation on the $(1\bar{1}1)$ plane oriented along the $[\bar{1}12]$ direction are shown in Figure~\ref{fig:ipn2d_Al_edge}. The corresponding distributions for a screw dislocation on the $(1\bar{1}1)$ plane with line direction $[110]$ are shown in Figure~\ref{fig:ipn2d_Al_screw}. The distributions in Figure~\ref{fig:ipn2d_Al_edge} and Figure~\ref{fig:ipn2d_Al_screw} are shown for different domain sizes and different mesh sizes to highlight the convergence of the numerical procedure for various discretization choices. The parameters used to obtain these figures are as follows: $L = 50.5b', 75.5b'$, $N_s = 101, 801, 1201$, $N_\tau = 202, 302$, $n_{AA} = 25$. Here, $b' = \sqrt{3}b$ for screw dislocations and $b' = b$ for edge dislocations on $\{111\}$ planes. The simulations for all these cases converge within a few hundred iterations, and hence are competitive in terms of computational efficiency with matrix based techniques like the SVPN method.
The dislocation density profile is shown in Figure~\ref{fig:ipn2d_Al_edge} and Figure~\ref{fig:ipn2d_Al_screw} for an edge and screw dislocation, respectively. It is seen that the edge dislocations splits while the screw dislocation has a wide core that may be thought of as two very closely spaced partials. These results agree with those present in \cite{schoeck01}, \cite{schoeck02}, \cite{schoeck12}. The separation of the partials in the case of edge dislocations is approximately 8.0\AA, which is in good agreement with experimental results as reported in \cite{schoeck02}. It is pointed out that unlike the SVPN method which predicts the splitting of the core for screw dislocations \cite{LKBK00}, the current method does not; this is in line with the predictions made in \cite{schoeck12}.

\begin{figure}[h!]
\centering
\includegraphics[width=0.7\textwidth]{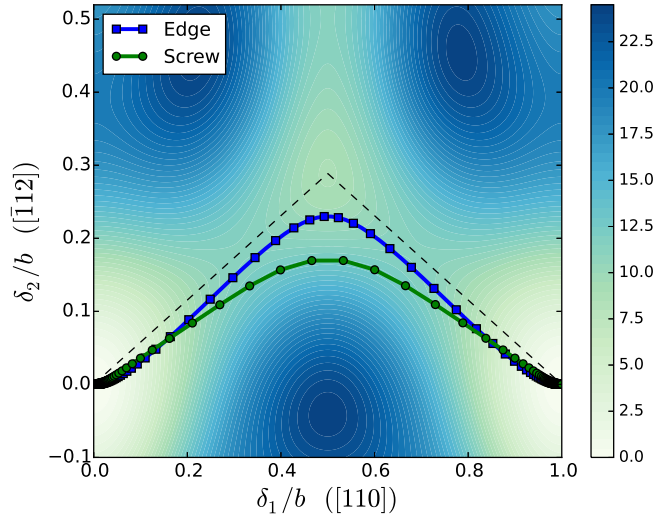}
\caption{Slip distribution of both edge and screw dislocations, oriented along $[\bar{1}12]$ and $[110]$ directions, respectively, on the $(1\bar{1}1)$ plane in Aluminium plotted over the generalized stacking fault energy contours. The theoretical dissociation path into Shockley partials is shown using black dashed lines.}
\label{fig:slip_gamma_Al}
\end{figure}

The slip distribution of both the edge and screw dislocations on the $(1\bar{1}1)$ plane is more conveniently visualized by plotting the components of slip along the $[110]$ and $[\bar{1}12]$ directions over the stacking fault energy surface, as shown in Figure~\ref{fig:slip_gamma_Al}. This clearly shows that the core of the dislocation is characterized by the splitting of the dislocation into Shockley partials, as expected. The iterative scheme proposed in this work for solving the generalized PN model is thus useful to study the core structure of dislocations in real materials, given appropriate generalized stacking fault energy data. 

\section{Discussion and Summary}
The results presented so far indicate that the iterative scheme proposed in this work is a reliable and computationally efficient alternative to solve the generalized Peierls-Nabarro model. It is emphasized that the matrix-free nature of the present approach allows for efficient parallelization of the iterative solution scheme. A number of issues pertinent to the iterative scheme proposed in this work to solve the generalized PN model and topics for future investigations are now discussed, followed by a summary of the key ideas presented in this work.   

\subsection{Handling the constant term in the misfit stress distribution}
It is to be noted that the constant term $d_0$ in the Fourier representation \eqref{eq:tau_slip_ft} of the misfit stress distribution $\tau_\alpha(s_1(x), s_2(x))$ on the slip plane is handled in the same manner as an external stress. Specifically, the misfit stress distribution is modeled as
\begin{displaymath}
\tau_\alpha(s_1(x), s_2(x)) = d_{\alpha,0} \chi_{[-L,L]}(x) + \sum_{n \neq 0, n = -\infty}^{\infty} d_{\alpha,n} \, \exp ink_0x.
\end{displaymath}
In the 1D case where the stacking fault energy is of the form \eqref{eq:gamma_sin_1d}, it is expected that $d_0 = 0$. This has been verified numerically.

\subsection{Choice of initial guess for slip distribution}
The convergence of the iterations depends on a good initial guess for the slip distribution. In particular, an initial guesses based on the approach developed in \cite{OP99} can be used, as is now illustrated in a 1D setting. Assuming that the stacking fault energy is piecewise quadratic of the form
\begin{displaymath}
\gamma(\delta) = \frac{1}{2}G(\delta - m)^2,
\end{displaymath}
where $G$ is a constant, $m$ is related to the amount of plastic slip, the inverse PN model reduces to the following equation
\begin{displaymath}
\frac{K}{2}\rho(x) = \frac{G}{\pi}\int_{-\infty}^{\infty} \frac{s(x') - m(x')}{x - x'}\, dx'.
\end{displaymath}
Assuming further that $s$ and $m$ admit a Fourier expansion of the form
\begin{displaymath}
s(x) = \sum f_n \exp ink_0x, \qquad m(x) = \sum g_n \exp ink_0x,
\end{displaymath}
where $k_0 = \pi/L$, it is straightforward to verify that
\begin{displaymath}
f_n = \frac{Gg_n}{Gg_n + \frac{1}{2}nk_0 \,\text{sgn}(n) K}.
\end{displaymath}
This approach has been used successfully to develop a phase field model for dislocation dynamics in \cite{KCO02}. Within the context of the iterative scheme proposed in this work, the foregoing solution can be used as an initial guess with $G$ computed from the curvature of the stacking fault energy at zero slip. While this provides a good initial guess for the iterative scheme proposed here, a simple piecewise constant slip distribution was found to be sufficient in practice.

\subsection{Computing the Peierls stress}
As remarked earlier, the current framework doesn't provide a direct means to compute the Peierls stress. A lower bound for the Peierls stress may, however, be obtained by estimated by considering a coarser mesh discretization. In the (unphysical) limit when the mesh size $h = \text{max}(h_j) \to 0$, the discretization gets closer to the continuum PN model, as illustrated by the numerical examples presented earlier. However, as is the case with the original PN model, information about the discreteness of the atomic lattice is lost in the limit $h \to 0$. A possible solution to introduce lattice discreteness directly into the current model is to make the choice $h_j = b$ (for every $j$) and consider the cases when $L/b$ is large. This is a physically meaningful choice since lattice misfits due to the presence of a dislocation are only truly defined at discrete locations corresponding to the perfect lattice points on the slip plane. This is similar to the procedure followed in the conventional PN model to compute the Peierls stress, where the energy functional is evaluated as a discrete sum of its values at lattice points, thereby introducing the necessary periodicity in the energy landscape. It is to be noted that the PN model is no longer valid in such a setting, and alternative approaches like the discrete dislocation equation \cite{wang15} that generalize the PN model to a corresponding discrete model may need to be used. Approximating the PN model directly with the coarse mesh whose size is equal to the lattice spacing thus introduces errors, as can be seen from Figure~\ref{fig:comp_pn1d_exact}, Figure~\ref{fig:conv_ipn1d_L}, Figure~\ref{fig:conv_ipn1d_xi} and Figure~\ref{fig:comp_pn1d_partials}. An estimate of the Peierls stress can be obtained by following a procedure similar to that adopted in the SVPN method \cite{BK97} as the external stress at which the generalized PN model fails to converge, or using the procedure describe in \cite{MGvSF01}. The solution of the inverse Peierls-Nabarro model in the presence of an external stress requires additional care since the domain size $L$ directly figures in the approximate model of external stress \eqref{eq:ext_stress_ipn}, and consequently in \eqref{eq:ipn_F}. Preliminary simulations showed domain size effects in the estimates of the Peierls stress. Alternatively, the approach adopted in \cite{MGvSF01} to compute the Peierls stress could be employed here. The continuum energy of a crystal with a dislocation is modified in \cite{MGvSF01} to include terms that reflect the periodicity of the lattice on the slip plane. This is accomplished by taking advantage of the specific ansatz introduced in \cite{Lejcek76} and using it in conjunction with the approach developed by Nabarro \cite{nabarro47} to estimate the Peierls stress. The use of the discrete Hilbert transform right from the start is another modeling choice that could prove to be helpful. These and related issues pertaining to the computation of the Peierls stress within the framework of the inverse Peierls-Nabarro model will be investigated in a future work. 

\subsection{Nonlocal misfit energy}
To overcome the limitations of the use of the generalized stacking fault energy to represent the misfit energy due to a specified slip distribution on the slip plane, nonlocal terms in the misfit energy have been introduced in \cite{MPBO98}. The present formulation motivates in a natural manner how such non-local terms may be introduced. To see this, the property of the inverse of the Hilbert transform \eqref{eq:ht_inv} and the property \eqref{eq:ht_const} are used to transform \eqref{eq:pn_hilbert} to get
\begin{displaymath}
\begin{split}
\sum_\beta \frac{1}{2} K_{\alpha\beta} \rho_\beta(x) &= -\mathcal{H}\left(\tau_\alpha(s_1(x), s_2(x))\right)\\
 &= -\frac{1}{\pi} \int_{-\infty}^{\infty} \frac{\tau(s_1(x'), s_2(x'))}{x - x'}\,dx',
\end{split}
\end{displaymath}
upto an additive constant that is determined by boundary conditions as before. The form of this equation suggests a natural non-local extension of the inverse PN model as follows:
\begin{displaymath}
\sum_\beta \frac{1}{2} K_{\alpha\beta} \rho_\beta(x) = -\frac{1}{\pi} \int_{-\infty}^{\infty} \frac{\tau_\alpha(s_1(x'), s_2(x'))}{x - x'}\,dx' + \frac{1}{\pi}\int_{-\infty}^{\infty} \frac{A_\alpha(s_1(x), s_2(x); s_1(x'), s_2(x'))}{x - x'} \, dx',
\end{displaymath}
where $A_\alpha(s_1(x), s_2(x); s_1(x'), s_2(x'))$ is a non-local term that can be modeled using techniques like those developed in \cite{MPBO98}. An iterative strategy for such non-local models can be developed in a manner analogous to the one presented in this work. These non-local extensions that extend the generalized PN model using data from lower scale models will be investigated in a future work.

\subsection{Extension to two dimensions}
The special properties of the Hilbert transform for functions of one variable have been exploited in the current work to develop the iterative scheme. To extend the current approach to two dimensions, as will be required for studying slip distributions on the slip plane, an extension of the Hilbert transform to handle the `$1/r$' kernel in \eqref{eq:gen_pn} in dimensions larger than one that simultaneously preserve all the \emph{nice} properties of the 1D Hilbert transform is desired. A variety of extensions of the Hilbert transform to higher dimensional equivalents have been proposed in the past: \cite{duffin57}, \cite{BdKdS06}. The extension of the iterative scheme developed in this work to the two-dimensional setting is also planned for a future work. It is worth reiterating, however, that the generalized PN model \eqref{eq:gen_pn} has been widely used to study the dislocation core structure in realistic materials despite the fact that it requires the slip distribution to be a function of one variable only. The iterative scheme developed in this work can thus be useful to study dislocation core structures in materials of practical interest.

\subsection{Conclusion}
To summarize, a novel numerical method that exploits certain properties of the Hilbert transform to solve the generalized Peierls-Nabarro model by reducing it to a fixed point iteration scheme has been developed in this work. The key advantage of the new method in comparison with standard approaches to solve the PN model is that it is matrix-free since it is of the form of a local fixed point iteration which is computationally efficient and is therefore amenable to parallelization schemes. The form of the inverse PN model further suggests in a natural manner extensions to non-local models that is compatible with the iterative structure developed here. The utility of the proposed method has been illustrated with a variety of examples. In particular, it is demonstrated that the proposed numerical scheme can handle realistic dislocation cores by illustrating its application to study the core structure of edge and screw dislocations in Aluminium on the close-packed $\{111\}$ planes. The dissociation of edge dislocations, the width of the stacking fault, and the diffuse core structure of screw dislocations that have been reported in previous works are captured satisfactorily by the present model. A possible strategy to include external stresses within the current framework and its limitations are discussed. The incorporation of external stresses in the inverse PN model so as to predict the Peierls stress and the change in the core structure of dislocations as it moves across the lattice will be carried out in a future work.  

\bibliographystyle{tfq}
\bibliography{inverse_pn_preprint}

\appendix

\section{Hilbert transforms} \label{app:hilbert_transform}
The Hilbert transform \cite{MJoh} of a function $f:\mathbb{R}\to\mathbb{R}$ is defined as the function $\mathcal{H}(f):\mathbb{R}\to\mathbb{R}$ given by
\begin{equation} \label{eq:def_hilbert_transf} 
\mathcal{H}(f)(x) = \text{P}\,\int_{-\infty}^{\infty} \frac{f(y)}{x - y}\,dy,
\end{equation}
where
\begin{displaymath}
\mathcal{H}(f)(x) = \lim_{\epsilon \to 0} \int_{x - \frac{1}{\epsilon}}^{x -\epsilon} \frac{f(y)}{x - y}\,dy + \int_{x + \epsilon}^{x + \frac{1}{\epsilon}} \frac{f(y)}{x - y}\,dy,
\end{displaymath}
is the Cauchy principal value of the improper integral in the Hilbert transform \eqref{eq:def_hilbert_transf}. The Hilbert transform is widely used for applications in signal processing. Some useful properties of the Hilbert transform that are relevant to this work are summarized here.
\begin{enumerate}
\item The inverse Hilbert transform of a given function is equal to the negative of the Hilbert transform of the function: if $\mathcal{H}(f)(x) = g(x)$, then $\mathcal{H}(g)(x) = -f(x)$. Thus
\begin{equation} \label{eq:ht_inv}
\mathcal{H}(\mathcal{H}(f))(x) = -f(x).
\end{equation} 
\item The Hilbert transform of a constant $k$ is zero:
\begin{equation} \label{eq:ht_const}
\mathcal{H}(k) = 0.
\end{equation} 
\item The Hilbert transform of complex exponentials have an especially simple form:
\begin{equation} \label{eq:ht_expi}
\mathcal{H}(\exp i\omega x) = -i \, \text{sgn}(\omega) \exp i\omega x,
\end{equation}
where $\text{sgn}(\omega)$ is the sign of $\omega$. Thus, the Hilbert transform \emph{rotates} a sinusoidal signal by $\pi/2$. As a particular case of the foregoing result, note that for any $\omega > 0$,
\begin{equation} \label{eq:ht_sin_cos}
\mathcal{H}(\sin \omega x) = -\cos \omega x, \qquad \mathcal{H}(cos \omega x) = \sin \omega x.
\end{equation}
\item If $\chi_{[a,b]}$ denotes the characteristic function for the set $[a,b]$, then 
\begin{equation} \label{eq:ht_characteristic_fn}
\mathcal{H}(\chi_{[a,b]})(x) = \frac{1}{\pi} \log \bigg\lvert \frac{x - a}{x - b}\bigg\rvert.
\end{equation}
Using \eqref{eq:ht_characteristic_fn} and \eqref{eq:ht_inv}, it is easy to see that
\begin{equation} \label{eq:ht_inv_characteristic_fn}
\chi_{[a,b]}(x) = \mathcal{H}\left(\frac{1}{\pi}\log \bigg\vert \frac{x - b}{x - a}\bigg\vert\right).
\end{equation}
\end{enumerate}
In the main text, $\mathcal{H}(f)(x)$ is often written as $\mathcal{H}(f(x))$.

\section{Acceleration schemes for fixed point iteration} \label{app:anderson_accn_fpi}

The convergence of the iterative scheme \eqref{eq:ipn_predictor_final} presented earlier to solve the Peierls-Nabarro model could be slow, as is typically the case with fixed point iteration strategies. A variety of acceleration schemes have been developed in the past to speed-up fixed point iterations. These are briefly outlined here in the context of solving the nonlinear fixed point equation
\begin{equation} \label{eq:nl_eqn_fpi}
f(x) = x,
\end{equation}
where $x \in \mathbb{R}^n$ and $f:\mathbb{R}^n \to \mathbb{R}^n$. The naive fixed point iteration, also known as Picard iteration, to find solutions of \eqref{eq:nl_eqn_fpi} takes the form
\begin{equation} \label{eq:fpi_scheme}
x^{(k + 1)} = f(x^{(k)}).
\end{equation}
A relaxed form the iterative scheme \eqref{eq:fpi_scheme} is obtained, for a given choice of $\lambda \in (0,1]$, as
\begin{equation} \label{eq:fpi_relax}
x^{(k + 1)} = (1 - \lambda) x^{(k)} + \lambda f(x^{(k)}).
\end{equation}
A useful means to rewrite \eqref{eq:fpi_relax} is as follows:
\begin{equation} \label{eq:fpi_ode_euler}
\frac{x^{(k + 1)} - x^{(k)}}{\lambda} = f(x^{(k)}) - x^{(k)},
\end{equation}
which can be thought of as the forward Euler discretization of the differential equation
\begin{equation} \label{eq:fpi_ode}
\dot{x} = g(x), \qquad g(x) = f(x) - x.
\end{equation}
Restrictions on the size of $\lambda$ thus correspond to well known restrictions on the choice of time step for the numerical solution of ordinary differential equations. Indeed, methods that accelerate the solution of the nonlinear equation \eqref{eq:fpi_ode} using specially designed Runge-Kutta methods have been proposed (see \cite{AE04}). The iterative scheme \eqref{eq:ipn_update} for the inverse Peierls-Nabarro equation \eqref{eq:ipn_predictor_final} proposed in this work is accelerated using an alternative strategy known as Anderson acceleration \cite{WN11}. The $n_{AA}$-stage Anderson acceleration for the nonlinear fixed point equation \eqref{eq:nl_eqn_fpi} is computed using $n_{AA}$ successive iterates of a fixed point iterative update, like \eqref{eq:fpi_scheme}. The basic idea behind the $n_{AA}$-stage Anderson acceleration scheme is to choose constants $\{\alpha_1, \ldots, \alpha_{n_{AA}}\}$ such that they minimize the residual
\begin{displaymath}
\left\lVert \sum_{i=1}^{n_{AA}} \alpha_{i} (f(x^{(k - i + 1)}) - x^{(k - i + 1)}) \right\rVert
\end{displaymath}
subject to the constraint 
\begin{displaymath}
\sum_{i=1}^{n_{AA}} \alpha_i = 1. 
\end{displaymath}
Thus, the current iterate is chosen as the particular linear combination of the $n_{AA}$ preceding iterates that minimizes the joint residual. Specific details regarding the numerical implementation of the foregoing minimization problem can be found in \cite{WN11}. For the present work, the first $n_{AA}$ iterations are carried out using the relaxed Picard iteration scheme \eqref{eq:fpi_relax} with a small valued of $\lambda$ to bootstrap the process. Subsequent iterations are computed using the $n_{AA}$-stage acceleration scheme as outlined above. 

\end{document}